\documentstyle[12pt]{article}

\newcommand{\bmat}{\left(\begin{array}}
\newcommand{\emat}{\end{array}\right)}
\def\NPB#1#2#3{Nucl. Phys. B{#1} (19#2) #3}
\def\PLB#1#2#3{Phys. Lett. B{#1} (19#2) #3}

\def\PRD#1#2#3{Phys. Rev. D{#1} (19#2) #3}
\def\PRL#1#2#3{Phys. Rev. Lett. {#1} (19#2) #3}

\def\yzero{\smash{\hbox{$y\kern-4pt\raise1pt\hbox{${}^\circ$}$}}}

\def\g{\gamma}

\def\beq{\begin{equation}}
\def\eeq{\end{equation}}
\def\beqa{\begin{eqnarray}}
\def\eeqa{\end{eqnarray}}
\def\Om{\Omega}

\def\-{\hphantom{-}}
\def\ov{\overline}
\def\s2{\frac{1}{\sqrt2}}

\def\oh{\frac{1}{2}}
\def\beq{\begin{equation}}
\def\eeq{\end{equation}}
\def\beqa{\begin{eqnarray}}
\def\eeqa{\end{eqnarray}}

\def\Tr{{\rm Tr \,}}
\def\diag{{\rm diag \,}}
\def\IF{\relax{\rm I\kern-.18em F}}
\def\II{\relax{\rm I\kern-.18em I}}
\def\IP{\relax{\rm I\kern-.18em P}}
\def\IC{\relax\hbox{\kern.25em$\inbar\kern-.3em{\rm C}$}}
\def\IR{\relax{\rm I\kern-.18em R}}

\def\cn{{\cal N}}

\def\cp{{\cal P}}

\def\NN{{\cal N}}
\def\Dsl{\,\raise.15ex\hbox{/}\mkern-13.5mu D} 
\def\IZ{Z\kern-.4em  Z}
\def\id{{\rm I}}
 \def\cp#1{\relax\ifmmode {\IP\kern-2pt{}_{#1}}\else $\IP\kern-2pt{}_{#1}$\=fi}
\newcommand{\drawsquare}[2]{\hbox{%
\rule{#2pt}{#1pt}\hskip-#2pt
\rule{#1pt}{#2pt}\hskip-#1pt
\rule[#1pt]{#1pt}{#2pt}}\rule[#1pt]{#2pt}{#2pt}\hskip-#2pt
\rule{#2pt}{#1pt}}

\newcommand{\Ysymm}{\raisebox{-.5pt}{\drawsquare{6.5}{0.4}}\hskip-0.4pt%
        \raisebox{-.5pt}{\drawsquare{6.5}{0.4}}}
\newcommand{\Yasymm}{\raisebox{-3.5pt}{\drawsquare{6.5}{0.4}}\hskip-6.9pt%
        \raisebox{3pt}{\drawsquare{6.5}{0.4}}}

\catcode`\@=11
\newdimen\@rotdimen
\newbox\@rotbox

\def\@vspec#1{\special{ps:#1}}
\def\@rotstart#1{\@vspec{gsave currentpoint currentpoint translate
   #1 neg exch neg exch translate}}
\def\@rotfinish{\@vspec{currentpoint grestore moveto}}
\def\@rotr#1{\@rotdimen=\ht#1\advance\@rotdimen by\dp#1%
   \hbox to\@rotdimen{\hskip\ht#1\vbox to\wd#1{\@rotstart{90 rotate}%
   \box#1\vss}\hss}\@rotfinish}
\def\@rotl#1{\@rotdimen=\ht#1\advance\@rotdimen by\dp#1%
   \hbox to\@rotdimen{\vbox to\wd#1{\vskip\wd#1\@rotstart{270 rotate}%
   \box#1\vss}\hss}\@rotfinish}%
\def\@rotu#1{\@rotdimen=\ht#1\advance\@rotdimen by\dp#1%
   \hbox to\wd#1{\hskip\wd#1\vbox to\@rotdimen{\vskip\@rotdimen
   \@rotstart{-1 dup scale}\box#1\vss}\hss}\@rotfinish}%
\def\@rotf#1{\hbox to\wd#1{\hskip\wd#1\@rotstart{-1 1 scale}%
   \box#1\hss}\@rotfinish}%
\def\rotate{\@ifnextchar[{\@rotate}{\@rotate[l]}}
\def\@rotate[#1]#2{\setbox\@rotbox=\hbox{#2}\@nameuse{@rot#1}\@rotbox}

\catcode`\@=12

\begin{document}

\makeatletter \@addtoreset{equation}{section} \makeatother
\renewcommand{\theequation}{\thesection.\arabic{equation}}
\pagestyle{empty}
\rightline{FTUAM-99/30; IFT-UAM/CSIC-99-37, DAMTP-1999-127,CAB-IB 2913799}
\rightline{\tt hep-th/9909172}
\vspace{0.5cm}
\begin{center}
\LARGE{\bf  
 Standard-like Models with Broken Supersymmetry
 from Type I String Vacua \\[10mm]}
\medskip
\large{G.~Aldazabal$^{1}$,
L.~E.~Ib\'a\~nez$^2$ and F. Quevedo$^3$
\\[2mm]}
\small{$^1$ Instituto Balseiro, CNEA, Centro At\'omico Bariloche,\\[-0.3em]
8400 S.C. de Bariloche, and CONICET, Argentina.\\[1mm]
$^2$ Departamento de F\'{\i}sica Te\'orica C-XI
and Instituto de F\'{\i}sica Te\'orica  C-XVI,\\[-0.3em]
Universidad Aut\'onoma de Madrid,
Cantoblanco, 28049 Madrid, Spain.\\[1mm]
$^3$ D.A.M.T.P., Silver Street, Cambridge, CB3 9EW England.
\\[6mm]}

\small{\bf Abstract} \\[7mm]
\end{center}

\begin{center}
\begin{minipage}[h]{14.5cm}
We   construct  $D=4$ Type I vacua with massless content
remarkably close to that of the standard model of particle physics.
 They are
tachyon-free non-supersymmetric models  which are obtained
 starting with a standard $D=4$, $N=1$ compact Type IIB orientifold and
adding the same number of Dp-branes and anti-Dp-branes distributed
 at different  points of the underlying orbifold.
Supersymmetry-breaking is felt by the observable world either
 directly, by  gravity mediation or gauge mediation,
depending on the brane configuration.
 We construct several simple three generation examples with
 the gauge group of the standard model or
 its left-right symmetric extensions.
The models contain a number of $U(1)$  gauge groups whose
anomalies are cancelled by a generalized Green-Schwarz mechanism.
These $U(1)$'s are broken but may survive as global symmetries
providing  for a flavour structure to the models.
 The value of the string scale
may be lowered down to the intermediate scale (as required in the
gravity mediation case) or down to 1-100 TeV for the  non-SUSY
models. Thus the present models are the first semirealistic
string vacua realizing the possibility of a low string scale.
 The unbalanced force
between the  pairs of Dp- and anti-Dp-branes provides for an
effect which tends to compactify some of the extra dimensions but
no others. This could provide a new mechanism for radius
stabilization.

\end{minipage}
\end{center}
\newpage
\setcounter{page}{1} \pagestyle{plain}
\renewcommand{\thefootnote}{\arabic{footnote}}
\setcounter{footnote}{0}

\section{Introduction}

During the past few years a beautiful picture has emerged
regarding the structure of string theory, in which all different
supersymmetric theories appear as different points of a general
moduli space. Understanding this moduli space is one of the big
open questions in this theory. However, one of the few things we
do know is that we do not live in such space, since supersymmetry
is not a symmetry of nature. Therefore it is mandatory to consider
the structure of non-supersymmetric strings.

Recently there has been some progress in understanding non
supersymmetric string models, by studying the dynamics of
brane-antibrane systems \cite{sen} and other constructions
\cite{others,ads,au}.
 This study has direct phenomenological interest due to the
observation that the fundamental string scale is not necessarily
of the order of the Planck mass and can be substantially lower
{\cite{witten,lykken,antoniadis}}.
 In particular type I strings may have very low
fundamental scale as long as the size of the extra dimensions is
large enough. Therefore we may in principle have a
nonsupersymmetric type I model with supersymmetry broken only at
the electroweak scale, something impossible to realize in the old
perturbative heterotic models.

The construction of four-dimensional models from type I theories
has been the subject of  research only recently
\cite{early,bl,ang,kak,fin,zwart,afiv,lpt,badagn} , since for many
years the activity concentrated mostly on heterotic strings.
Standard-like models from heterotic strings have been constructed
since several years already \cite{review,recent,ovrut}. Despite the remarkable similarity
with the Standard Model of particle physics, they all suffer from
different problems. In particular the standard
questions of supersymmetry breaking and moduli stabilisation could
not be approached from a string theory point of view. On the other
hand, supersymmetric
 type I model building has appeared  to be very restrictive
and no quasi-realistic models have emerged to date
\cite{real} . This
difficulty is related to the conditions coming from the
cancellation of Ramond-Ramond(RR) tadpoles which are extremely
restrictive.

In spite of recent interest in D-brane scenarios with a lowered
string scale
\cite{untev,antoniadis,bajogut,sundrum,
shiutye,bachas,kakutye,benakli,biq,imr,dr,rs},
 the studies presented in the literature up to now
have used the D-brane techniques  more as an inspiration than as
an explicit tool to construct new phenomenologically interesting
string vacua. A number of phenomenological issues
 have been discussed but never based on specific
semirealistic string vacua with D-branes. One of the purposes of
the present article is to fill this gap and provide specific
consistent string vacua in which these  issues may  be addressed.

We present  the first standard-like models explicitly
 constructed from
type I strings. This is a new class of vacua based on $\cn=1$
toroidal Type IIB
 orientifolds plus the inclusion of branes and anti-branes, generally
 stuck at different orbifold fixed  points
\cite{au}. The presence of these brane anti-brane pairs in the
vacuum makes substantially easier the construction of tadpole-free
models. Supersymmetry is only broken in some subsector of the
theory by the presence of anti-Dp-branes which  break
supersymmetry. Since these anti-Dp-branes may be isolated in the
bulk of compact space
 the $N=1$ sector of the theory may feel SUSY-breaking only
in a suppressed manner. We then have explicit string realizations
of
 the gravity mediated scenario. In other models the
spectrum is non-supersymmetric or else feels SUSY-breaking by
gauge mediation.
 Our models have three families of quarks and leptons
and are remarkably simple. We will present explicit examples with
the Standard Model gauge group and also left-right symmetric
models as
 well as models of the Pati-Salam type.

The structure of this paper is as follows. In the next section we
give an overview of the orientifold constructions of
four-dimensional chiral type I strings, including the recent
development of reference \cite{au}, where branes and anti-branes
are included in a consistent manner, cancelling all tadpoles and
projecting out tachyonic states. In section 3 we perform a general
treatment of the non-supersymmetric models in the presence of one
Wilson line which serves as the basis for our construction. We
provide the general massless spectrum for this class of theories
and show how $U(1)$ anomalies are cancelled via a generalized
Green-Schwarz mechanism along the lines of ref. \cite{sagnotti,iru}.

 In section  4.1  we present explicit
examples of quasi-realistic three-generation
models where only branes are stuck at
fixed points whereas anti-branes live in the bulk. Since the
anti-branes break supersymmetry, these are explicit examples of
gravity mediated supersymmetry breaking, which then ask for a
fundamental string scale to be an intermediate scale \cite{biq}.
Other example of gravity mediated models in which both branes and
anti-branes are stuck at fixed points is provided in section 4.3.
 Section  4.2 deals with
models with branes stuck at some points and anti-branes stuck at
different points. There are fields in these models carrying
quantum numbers of the anti-branes which can mediate supersymmetry
breaking either directly or through gauge interactions. In this
case we would need the fundamental scale to be as low as the TeV
scale \cite{lykken,untev,antoniadis}.

In chapter 5 we study a number of generic dynamical issues in this
class of models. We show how the string scale can be lowered down
to the intermediate scale (as required in the gravity mediated
models of sections 4.1 and 4.3) or below (as required in the
non-SUSY models of section 4.2). We discuss the presence of
attractive forces between branes and anti-branes as well as other
effects  which may lead to a potential for the moduli
\cite{au} . We also
study the general structure of Yukawa couplings and
 show how gauge couplings may naturally unify at the intermediate
scale in some particular
gravity mediated models of the class discussed in section
4.1.

Chapter 6 is left for our final conclusions and comments. The
appendices provide for  additional models which exemplify some of
the issues discussed in the text.

\section{Type IIB orientifolds in the presence of anti-Dp-branes}

In this section we introduce our notation and projection rules for
obtaining the massless spectrum in $D=4$ dimensions. We also
comment on different alternatives for building stable
non-supersymmetric vacua. We mainly follow references \cite{afiv} and  \cite{au}.

A Type IIB orientifold in four dimensions is obtained when such
theory,  toroidally compactified on $T^{6}$, is divided out by the
joint action of world sheet parity symmetry $\Omega$, exchanging
left and right movers, and a discrete symmetry group $G_1$ (like
$\IZ_N$ or $\IZ_N\times \IZ_M$), acting crystalographically on $T^6$.
The $\Omega $ action can be also accompanied by extra operations thus
leading to a general orientifold group $G_1+ {\Omega} G_2$ with
${\Omega}h {\Omega} h' \in G_1$ for $h,h' \in G_2$ In what follows
we mainly consider $\IZ_N$ twists. Twist eigenvalues, associated to
complex compactified  coordinates $X_a$ ($a=1,\dots, 3$), are
encoded into the vector $v={\frac{1}N(\ell_1,\ell_2,\ell_3)}$
where  $\ell_a$ are integers. Such integers are specified by the
number of required unbroken supersymmetries on each movers sector.
We choose them to satisfy $\ell_1+\ell_2+\ell_3=0$. A classification
for the $\cn=1$ case may be found in \cite{dhvw}. Orientifolding
closed Type IIB string introduces a Klein-bottle unoriented world-sheet.
Amplitudes on such a surface contain tadpole divergences. Tadpoles
may be generically interpreted as unbalanced orientifold planes
charges under  RR form potentials.
 Such unphysical divergences can be eliminated by introducing Dp-branes,
 carrying opposite charges.

 In this way, divergences occurring in the open string sector
cancel  up the closed sector ones and produce a consistent theory.
Moreover, sets of branes and antibranes can be consistently
introduced in order to cancel RR tadpoles. Even though  the
models are non-supersymmetric (due to the presence of the
anti-D-branes), it has been shown in \cite{au} that 
 models free of tachyons can
be easily constructed. Since, in the orientifolds we are
considering, supersymmetry is preserved in the closed string
sector, and is only broken in the open string sector by the
presence of the anti-D-branes, most of the usual techniques in
supersymmetric orientifolds hold
\cite{orient,bl,ang,kak,afiv}. Hence, we
focus on the open string sector. Open string states are denoted by
$|\Psi, ab \rangle \lambda ^{pq} _{ab}$. Here, $\Psi$ refers to
world-sheet degrees of freedom whereas $a,b$ are Chan-Paton
indices associated to  the open string endpoints lying on
D$p$-branes and D$q$-branes respectively. $\lambda ^{pq}$ is the
Chan-Paton, hermitian matrix, containing the gauge group structure
information. Analogously, $\lambda ^{\ov p \ov q}$ ($\lambda ^{\ov
p q}$) is introduced for open strings ending at D$\ov p$, D$\ov
q$-antibranes ( D$\ov p$ antibrane,  D$ q$-brane, etc.). We will
denote these situations generically by introducing capital indices
$P=p,\ov p$ and $Q=q,\ov q$.
 The action of an element of the
orientifold group on Chan-Paton factors is achieved by a unitary
matrix $\gamma _{g,P}$ such that $g: \lambda ^{PQ} \rightarrow
\gamma _{g,P} \lambda ^{PQ} \gamma^{-1}_{g,Q}$. We denote the
matrices  associated to the $\IZ_N$ orbifold twist $\theta ^k $ and
$\Om$, acting on a P-brane, as $\gamma _{k,P}$ and  $\g_{\Om,P }$
respectively. Consistency under group transformations imposes
restrictions on the twist matrices  $\gamma _{g,P}$ (see for
instance \cite{afiv}). A consistent choice for world sheet
parity action representation, which we follow below is
\beqa
\g_{\Om,9 } & = & \g^T_{\Om,9 } \nonumber \\ \g_{\Om,5 } & =&
-\g^T_{\Om,5} \label{gpa}
\eeqa
Also, generic matrices for the $\IZ_N$
orbifold twist, with $N=2M$ ($N=2M+1$), can be provided. They
can be defined as
\beq \gamma_{1,P}=({\tilde
\gamma_{1,P}},{\tilde \gamma _{1,P}}^{*})
\label{gtilde}
\eeq
with $* $ denoting
complex conjugation and where  ${\tilde \gamma }$ is a $N_P\times
N_P$ diagonal matrix given by
\beq {\tilde \gamma}_{1,P}   = \diag
(\cdots,\alpha^{NV_j}I_{n_j^p},\cdots, \alpha^{NV_ M} I_{n_M^P})
\label{gp} \eeq
with $\alpha = {\rm e}^{2i\pi /N}$ and
$2N_P=\sum_{j=1}^M  n_j^P$ is the number of P-branes. The choice
$V_j=\frac{j}N$ with $j=0,\dots, M$ corresponds to an action
``with vector structure '' ($\gamma _{1,P} ^N=1$) while
$V_j=\frac{2j-1}{2N}$ with $j=1,\dots,M$ describes an action
``without vector structure" ($\gamma_{1,P}
^N=-1$)\footnote{Following the classification introduced in
\cite{blpssw} for six-dimensional models.}.
 In order to  compute the spectrum, it proves useful to associate a
``shift vector" $V^P$, with $V_j$ coordinates, to the twist matrix
${\gamma}_{1,P}$. Namely, \beq {\gamma}_{1,P} \rightarrow V^{P}
= ( V_1 ,\dots ,V_M) \label{vp} \eeq
 The models discussed below are   mainly
based on  $\IZ_3$ orientifolds. The twist matrix and shift vector in such
case are
\beq {\tilde \gamma}_{1,P}   = \diag
(I_{n_0^p}\cdots,\alpha I_{n_1^p}) \rightarrow V^{P}   =
\frac{1}3(0\dots 0\,1,\dots 1)= \frac{1}{3}\
 (0^{n_0^p}, 1^{n_1^p})
\eeq
with $\alpha = {\rm
e}^{2i\pi /3}$. As usual the powers of the entries in the vector
$V^P$ refer to the number of times each entry appears.
 By choosing $\g_{\Om,9}$ and $\g_{\Om,5}$ matrices
 \beq
 \g_{\Om,9} =
\bmat{cc}0&\id_{N_9}\\ \id_{N_9} & 0 \emat \quad ; \quad
\g_{\Om,5} = \bmat{cc}0&-i\id_{N_5}\\ i\id_{N_5} & 0 \emat
\label{goms95} \eeq
the  consistency constraint (associated to group operation
${(\Om\theta ^k)}^2 =\theta ^{2k}$) \beq \g_{k,P}^* =  \g_{\Om,P}
\, \g_{k,P} \g_{\Om,P} \label{famp} \eeq is satisfied with the twist
matrices defined in (\ref{gp}).
 The requirement of tadpoles cancellation leads to further
 restrictions on twist matrices. Such constraints  have been extensively
discussed in the literature for supersymmetric orientifolds. It
was shown in \cite{au} that essentially the  same constraints are
valid, when antibranes are present, if $\Tr\g_{k,p}$ is replaced
by $ (\Tr\g_{k,p}-\Tr\g_{k,\ov p})$ in the tadpole cancellation
equations of the supersymmetric case and the added number of branes
and anti-branes is the same. We refer to this point
when we  construct specific models.

\subsection{ The massless spectrum}
The massless spectrum in an array of
branes and antibranes was discussed in \cite{ads,au}. Here we follow 
the discussion in Ref. \cite{au} generalizing it for the case of
nonvanishing Wilson lines. In the case of
supersymmetric orientifolds, it  was shown in Ref. \cite{afiv}
that computation of spectra is greatly simplified if a
Cartan-Weyl basis is chosen. Let us recall the basic ingredients
of such a construction and indicate how this is easily generalized
to uncover the description of systems containing both branes and
antibranes. In a Cartan-Weyl basis, Chan-Paton generators are
organized into charged generators $\lambda_a = E_a$, $a=1,\cdots,
{\rm dim}\, G_P - {\rm rank}\, G_P$, and Cartan algebra generators
$\lambda_I = H_I$, $I=1,\cdots, {\rm rank}\, G_P$, where $G_P$ is
the gauge group on the P-brane. If a vector $H=(H_1,\dots,H_{{\rm
rank}\, G_P})$ is defined, then
\begin{equation}
[H, E_a]=\rho^aE_a \label{cw}
\end{equation}
where the (${\rm rank}\, G_P$)-dimensional vector  $\rho ^a$ is
the root vector associated to the generator $E_a$.
 Matrices  $\gamma_{1,p}$ (and their powers) represent the action of the
$\IZ_N$ group on Chan-Paton factors, and they correspond to elements
of a discrete subgroup of the Abelian group spanned by the Cartan
generators. Thus they can be written as
\begin{equation}
\gamma _{1,p}= e^{-2i\pi V^p \cdot H } \label{Vdef} \eeq
Generically, invariance of open string massless states under the
orientifold group action in $PP$ sectors leads to constrains on
Chan-Paton matrices of the form
\beqa \lambda  & = &  \pm
\gamma_{\Omega,P} \ {\lambda ^T} \ \gamma_{\Omega,P}^{-1}\nonumber\\
\lambda & = & e^{2\pi i {\frac{k}N}} \gamma_{\theta,P} \ \lambda \
\gamma_{\theta,P}^{-1} \label{projg} \eeqa

The first equation,
imposed by $\Om $ orientifold projection, performs a first
selection of allowed $\rho ^P $ weight vectors. The second one,
required for invariance under orbifold twists, further projects
such vectors through the simple constraint
${\rho}^P.V^P={\frac{k}N}  \quad  {\rm \, mod \,} {\IZ} $.
For instance, in ${\bf 55}$ sectors the projections for
D5$_3$-branes at fixed points  are given by \beqa
\begin{array}{lll}
\lambda^{(0)} = \gamma_{\theta,5}\ \lambda^{(0)}\
\gamma_{\theta,5}^{-1} & \quad & \lambda^{(0)} = -
\gamma_{\Omega,5}\ {\lambda^{(0)}}^T\ \gamma_{\Omega,5}^{-1}\\
\lambda^{(i)} = e^{2\pi i v_a}\ \gamma_{\theta,5}\ \lambda^{(i)}\
\gamma_{\theta,5}^{-1} & \quad & \lambda^{(i)} = \pm
\gamma_{\Omega,5}\ {\lambda^{(i)}}^T\ \gamma_{\Omega,5}^{-1}
\label{proj2}
\end{array}
\eeqa with positive sign for $a=1,2$ and negative for $a=3$.
 The projections
are the same for both bosonic and fermionic states, thus leading
to $\NN=1$ supermultiplets. The $\Om$ projections, with $\g_{\Om,5
}$ defined in (\ref{goms95}), select the $Sp(2N_5)$ root vectors
$\rho  _S= ({\underline \pm 1,\pm 1,0\dots0})$ and
$\rho_L=({\underline \pm 2,0\dots0})$ (and Cartan generators) when the  minus sign in
the first equation in (\ref{projg})  is present. If there is instead
a plus sign, long root vectors are absent. Hence, $\lambda^{(0)}$ constraint on
the left
\begin{equation}
\rho^a \cdot V^P= 0 {\rm \, mod \,} {\IZ} \label{55gg}
\end{equation}
selects the gauge group, a subgroup of $Sp(2N_5)$  whereas matter
states correspond to charged generators with
\begin{equation}
\rho^a \cdot V^P= v_a {\rm \, mod \,} {\IZ} \label{m9p}
\end{equation}
With $v_a=\ell_a/N, a=1,2,3$, the components of the vector that defines
the orbifold twist. The adequate $\rho $ vectors must be used in
each complex direction.

Recall that in supersymmetric orientifolds with
D5-branes (thus with  even $\IZ_N$ twists) at fixed points,
tadpole equations generically require  shift vectors 
``without vector structure" and thus, long root vectors do not
contribute to the massless spectrum \cite{afiv}. However, the
situation is different when antibranes are also present. In this
case even twists ``with vector structure" or odd twist with
D5-branes and D$\ov 5$-antibranes can also appear. For instance,
long root vectors complete symmetric representations  of unitary
groups, not present in supersymmetric cases.

Similar considerations are valid when dealing with ${\bf \ov 5 \ov 5}$ sectors.
Exactly the same projections apply to  NS states  while an extra minus
sign appears for R states in the $\Om $ projections. In  mixed
sectors, ${\bf PQ}$ ( ${\bf PQ}= {\bf 59} ,{\ov {\bf 5}}{\bf
9},{\ov {\bf 5}}{\bf 9}$ etc.), with string endpoints lying on
different kinds of branes,  the subset of roots of $G_P\times
G_Q$ of the form
\beq \rho _{PQ}= (W_{P}; W_{Q})=
({\underline {\pm 1, 0,  \cdots, 0}};{ \underline {\pm 1, 0,
\cdots, 0}}) \label{w95def} \eeq
must be considered. In these
cases, orientifold projection maps ${\bf PQ}$ into
$ {\bf QP} $ sector and, thus, it imposes no restrictions on the
spectrum. The relevant phases for NS and R states in each sector,
due to orbifold action, can be read from \cite{au}. Let us
consider, for instance, projections in ${\bf 59}$ and ${\bf
\ov{5}9}$ sectors. Recall that NS states are labeled by $|s_1,s_2
\rangle$ $SO(4)$ spinor weight  while R states correspond to
$|s_0;s_3 \rangle$ spinor weights ($s_j= \pm \oh $) where $s_0$ defines space-time
chirality . GSO projection demands $s_1=s_2$ and
$s_0=s_3$ in  ${\bf 59}$ sector, whereas opposite signs are needed
in ${\bf \ov{5}9}$ sector.
 For both, bosons   and fermions  in  ${\bf 59}$ sector we must have
\begin{equation}
\rho_{(95)} \cdot V ^{(95)}= \pm \oh(v_1 +v_2) {\rm \, mod \,}
{\IZ} \label{95}
\end{equation}
with  positive and negative  signs corresponding to $s_1=s_2=\pm
\oh$ and  $s_0=s_3=\pm \oh $  respectively.
For  NS states in
${\bf \ov{5} 9}$ sector we have,
\beq \rho_{(9 \ov 5)} \cdot V^{(95)}=  \pm \oh(v_1-v_2) {\rm \, mod \,} {\IZ}
\eeq
where positive and negative signs correspond to $s_1=-s_2=\pm\oh$.
Finally, orbifold projection on  R states requires
\beq \rho_{(9 \ov 5)}
\cdot V ^{(9\ov 5)}= \pm \oh v_3 {\rm \, mod \,} {\IZ}
\label{5b9} \eeq
with plus and minus signs corresponding to
$s_0=-s_3 = \mp\oh$. Notice that this last condition is similar to
(\ref{95}) but there the plus sign corresponds to $s_0=\oh$, i.e.
positive chirality fermions, while here it describes negative
chirality ones (we are using $v_1 +v_2+v_3=0$).

\subsection{Tachyon free non-supersymmetric models}
Generic models with branes and antibranes will contain tachyonic
states whenever  brane-antibrane pairs of the same type coincide.
More precisely, tachyons can be  present in ${\bf 9 \ov 9}$
sectors or  ${\bf 5}_L {\ov {\bf 5}}_L$ sectors, with 5D-branes
and $\ov 5$D-antibranes at the same fixed point $L$  and whenever
their Chan-Paton twist matrices do overlap \cite{au}.

Once the origin of tachyonic fields is identified different ways
for building tachyon free theories open up \cite{au}. Thus, models
with branes of only one type (say D9-branes)  and antibranes of
the other ($\ov 5$-antibrane) are automatically tachyon free.
Another  possibility  is to consider non-overlapping Chan-Paton
twist matrices for the coincident  branes and the antibranes. A
third way is to place branes and antibranes at different fixed
points (or have different Wilson lines, in a T-dual description)
of the internal space. It is this last route the one we will
mainly  pursue in building our models so let  us further  comment
on it.

On the one hand such kind of models indicates an attractive
possibility for breaking supersymmetry.  Namely,
non-supersymmetric antibrane sectors (at some fixed points) could
transmit supersymmetry breaking to the supersymmetric brane
sector (at different fixed points) through the exchange of bulk
fields. However, the situation is more subtle.
 If branes (or antibranes) are not really stuck at the fixed point, they
 will be able to move
through the bulk as a dynamical brane. Since brane-antibrane
systems develop a net attractive force between them, they will
eventually come to the top of each other and annihilate into the
vacuum. Thus, if  brane-antibrane annihilation is completely
successful and all antibranes disappear, we will end up with a
stable supersymmetric vacuum. In such a case, non-supersymmetric
models should be viewed as excitations of a supersymmetric vacuum.
Interestingly enough there are situations in which, due to  local
obstructions, only partial, or not  annihilation at all,  is
allowed. These appear as true stable vacua with both,
supersymmetric and non supersymmetric sectors. In the following
sections we will use two different ways for trapping  branes:
\bigskip

\noindent {\bf i. Through tadpole conditions:}

Whenever  orientifold planes RR twisted fields charges are not
vanishing, branes must be locally introduced  in order to cancel
them. Depending on the specific form of twist matrices, antibranes
are also needed for such a cancellation. Even though different
configurations of branes and antibranes might be possible,
antibranes cannot completely annihilate. Let us briefly illustrate
the idea in the $\IZ_3$ orientifold example which we further develop
in next section. In this case, tadpole cancellation requires
\cite{au}

\beq \Tr \gamma_{\theta,9} + 3(\Tr\gamma_{\theta,5,L}-
\Tr\gamma_{\theta,\bar{5},L}) = -4 \label{tadtrap} \eeq
where $L=1,\dots, 9$ denotes the nine orbifold fixed points
in the first two complex planes. Also the
number of fivebranes and antibranes must be the same. Notice that
absence of antibranes implies that neither 5D-branes are present
and the supersymmetric orientifold condition \cite{ang,kak} $\Tr
\gamma_{\theta,9}=-4$ is recovered. Notice instead that choosing
$\Tr \gamma_{\theta,9} \ne-4 $ inevitably demands the presence of
branes and/or antibranes at all fixed points
\footnote{In the presence of discrete Wilson lines on $9$-branes
the different fixed points feel different gauge embeddings. In this
case some of the fixed points may  not require the presence of $5$-branes
or anti-$5$-branes, as we  describe below.}
 As mentioned,
different arrays may be possible. In some situations adding
D5-branes (and no anti-$D5$-branes) at fixed points may be enough to
 achieve the twisted tadpole cancellation
above. Nevertheless, since the same number of antibranes is
required, they must be present in the bulk. Also configurations of
branes and antibranes distributed among the fixed points are possible.
Examples of such situations are provided in section 4. Recall
that, whereas in the supersymmetric case  the tadpole
conditions  have  a unique solution, a variety of non-supersymmetric
solutions are  possible.
\bigskip

\noindent {\bf ii. By sitting isolated branes at orientifold, but
not orbifold fixed  points.}

Assume that $\IZ_3$ orbifold twist generators  and $\Om R$
orientifold action, where $R$ is a $\IZ_2$ generator,  are present
in the orientifold group. This is the case for instance of $\IZ_6$
orientifold, where $R$ is a reflection in complex directions
$(X_1,X_2)$. This is also realized in the  T-dual version of $\IZ_3$
with 3 and 7D-branes where we have the element $(-1)^F R_1R_2R_3$,
with $R_i$ a reflection in $Y_i$  plane. In such a case we could
place an isolated antibrane at a $\IZ_2$ fixed point, which is not a
$\IZ_3$ fixed orbifold point. Orbifold invariance requires other two
antibranes at each of the $\IZ_3$ images. This antibrane triplet is
trapped since it would need a $\Om R$ mirror
 to leave the orientifold point as dynamical antibrane.
Hence, this triplet is  a non supersymmetric, bounded
configuration, of three antibranes which is stable. Building up a
full consistent stable non supersymmetric orientifold model will
require that branes are also stuck, otherwise they will be
attracted to the antibranes and annihilate. In fact, this appears to
be the case for $\IZ_6$ orientifold. In section 4.3 we provide a
$\IZ_3$ example with 3-branes and 3-antibranes. In principle, the
above construction could be extended to other orbifold actions
like $\IZ_3\times \IZ_3$, $\IZ_7$ etc.

\section{Type I  $\IZ_3$  vacua with Wilson lines}

The introduction of Wilson lines \cite{inq}\ in heterotic orbifold
\cite{dhvw}\ models enormously increased the possible consistent
chiral string vacua and allowed for  the first explicit construction of
quasi-realistic string models in the past decade \cite{iknq},
including standard-like models with three families of quarks and
leptons. The most interesting models, from the phenomenological
point of view, happened to be those built from the $\IZ_3$
orbifold. One of the reasons for this is that three generations
come out very naturally in this construction.

In this section, following \cite{afiv}, we will introduce the
effects of Wilson lines on the $\IZ_3$ Type IIB orientifolds
including branes and anti-branes as in ref. \cite{au}. The $\IZ_3$
orientifold without Wilson lines was already presented in
\cite{au}. We will see that  Wilson lines substantially increases
the versatility of the models, allowing to break the gauge groups
in the 9-brane sector and also the possibility to modulate and
reduce substantially the matter content from  the 5-brane sectors.

Adding one single Wilson line will give us enough freedom to
construct phenomenologically interesting three generation models,
and the spectrum is simple enough that we can discuss the most
general case explicitly.

\subsection{Spectrum of the Models}

Let us consider the twist matrix
$\gamma_{1,9}=({\tilde \gamma_{1,9}},{\tilde \gamma _{1,9}}^{*})$ as given
in \ref{gtilde}. We can also include a Wilson line
${ {\cal W}}=(\tilde {\cal W}, {\tilde {\cal W}_1}^*)$ wrapping
along direction $e_1$ in the first complex plane. The most general
 explicit form for such matrices can be written as
\beqa
{\tilde \gamma
_{9}}  & = & \diag (I_{N_0},I_{N_1},\alpha I_{N_2},\alpha
I_{N_3},\alpha I_{N_4}) \\ {\tilde {\cal W}} & = & \diag
(I_{N_0},\alpha I_{N_1}, I_{N_2},\alpha I_{N_3},\alpha ^2 I_{N_4})
\eeqa
 with $\alpha ={\rm
e}^{2i\pi /3}$.

The  associated shifts are
\beqa
 V_9 &=&
{\footnotesize{\frac{1}{3}}}\
(0^{N_0};0^{N_1};1^{N_2};1^{N_3};1^{N_4})\\ W_9 &=&
{\footnotesize{\frac{1}{3}}}
(0^{N_0};1^{N_1};0^{N_2};1^{N_3};2^{N_4}) \eeqa
where, as usual, the power on each of the entries actually means
the number of times that number appears in the vector. So each
vector has dimension $N_0+N_1+N_2+N_3+N_4= 16$.
 Hence the 99 sector gauge group is
 \beq
SO(2N_0)\times \prod _{s=1}^{4} U(N_s)
 \eeq
R-R tadpole cancellation requires
\beq
\Tr ({\cal W})^k \gamma_{\theta,9} +
3(\Tr\gamma_{\theta,5,L}- \Tr\gamma_{\theta,\bar{5},L}) = -4
\label{tadpzz} \eeq
for $k=0,1,2$. Also  the total number of branes and antibranes
must be the same.

In the first and second complex planes there are nine orbifold
fixed
 points which we label as $(a,i)$, $a,i=0, 1, 2$. We can put several
 5-branes
or anti 5-branes at each point as long as the tadpole conditions
are satisfied. In the general case we
 can choose to sit  $2n_0^{i}+2m_0^{i}$ 5-branes
 and $2p_0^j+2q_0^j$  $j\neq i$ anti 5-branes at
 the fixed points set
$L_0=\{(0,0),(0,1),(0,2)\}$, which do not feel the action of $\cal
W$.
 Here the indices
 $i,j=0,1,2$ label  each of the three fixed points, and the requirement
$i\neq j$ states that we do not put branes and anti-branes at the
same
 point. Similarly, to  the
fixed points  $L_1= \{(1,0),(1,1),(1,2)\}$ feeling $V+W$ we assign
$2n_1^i+2m_1^i$ 5-branes and $2p_1^j+2q_1^j$ anti 5-branes.
Finally
 at $L_2=\{(2,0), (2,1),(2,2)\}$ feeling $V-W$ we assign
 $2n_2^i+2m_2^i$
5-branes and $2p_2^j+2q_2^j$ anti 5-branes. We will also consider
 the possibility of having $2r$ extra anti 5-branes not attached to any of
 the fixed points.

Therefore at the  $ith$ fixed point in the set $L_a$, $a=0,1,2$ we
have:
\beqa
\gamma _{5,i,a} &   = & \diag  (I_{2m_{a}^i}, \alpha
I_{n_a^i},\alpha^2 I_{n_a^i}) \qquad \Tr\gamma_{5,i,a}  =  2\
m_a^i\ -\ n_a^i\nonumber\\ \gamma_{\ov 5,j,a} &  = &  \diag
(I_{2p_a^j}, \alpha I_{q_a^j},\alpha^2
 I_{q_a^j})\qquad \Tr\gamma_{\ov 5,j,a}= 2\ p_a^j-q_a^j
\eeqa

Since the number of branes and antibranes must be the same we must
have
\beq
\sum_{i,a}(n_a^i+m_a^i)=\sum_{b,j}(p_b^j+q_b^j) + r. \eeq

Using the explicit expressions for twist matrices, and
\beqa
 V_9+W_9 & = & {\footnotesize \frac{1}{3}}\
\left(0^{N_0};1^{N_1};1^{N_2};2^{N_3}; 0^{N_4}\right)\nonumber\\
V_9-W_9 & = & \frac{1}{3}\ \left(0^{N_0};2^{N_1};1^{N_2};0^{N_3};
2^{N_4}\right) \eeqa
the  tadpole equations \ref{tadpzz} read ($j\neq i$):
\beqa
n_0^i-2m_0^i &=& 2p_0^j- q_0^j\ =\ 12- N_2-N_3-N_4 \nonumber \\
n_1^i-2m_1^i &=& 2p_1^j- q_1^j\ =\ 12-N_1-N_2-N_3\nonumber \\
n_2^i-2m_2^i &=& 2p_2^j- q_2^j\ =\ 12-N_1-N_2-N_4
\label{tad} \eeqa

where we have used that $N_0+N_1+N_2+N_3+N_4 =16$.

The total  gauge group is (when all branes are at fixed points) thus
\beq
SO(2N_0)\times\prod _{s=1}^4 U(N_s) \times \prod_{a,i,j\neq
i=0}^2
 [Sp(2m_a^i)\times U(n_a^i)]\times [Sp(2p_a^j)\times
U(q_a^j)]
 \eeq
Following the method of the previous section we can compute
 the fermionic spectrum, which is supersymmetric on the branes and
 non-supersymmetric on the anti-branes.
First, the {\bf 99} sector is independent of the brane/anti-brane
distribution, and it is quite analogous to what one finds in
computing  the untwisted sector of heterotic orbifold models. In
particular, for the $\IZ_3$ orbifold, it comes in three identical
copies, associated to each of the three complex planes. This is
the origin of the three families in most of our models
\footnote{Since all these  sectors are supersymmetric there are
also massless complex scalar partners transforming exactly in the
same way.}
\beqa
{\bf 99} \qquad\quad {\rm Fermions_+}:\nonumber\\ & &   3\, [(\bf
{2N}_0,\bf {N}_2) + (\ov {\bf N}_1,{\bf N_3}) +({{\bf N}_1},{\bf
N}_4)+ ({\ov {\bf N}_3},\ov{\bf N}_4)+ \ov {\bf
a}_2]
 \nonumber
\label{99spec}
\eeqa
Where we write in parentheses pairs of fundamentals of the
corresponding $U(N_s)$ groups and $\ov {\bf a}_2$ is the
(conjugate) antisymmetric representation of $U(N_2)$. The {\bf 59}
sectors depend on which shift vector is felt by the corresponding
fixed points. Then, for the set $L_0$, the effective shift for the
$59$ sector is simply $V_9\times (0^{m_0^i};1^{n_0^i})$. We then
find
\beqa
 {\bf 5_{L_0}9} \quad {\rm Fermions_+} :\nonumber \\
& &
({\bf 2N}_0,{\bf n}_0^i) +
({\ov {\bf N}}_1,{\bf n}_0^i)+
 ( {\bf N}_1,{\bf n}_0^i ) +
({\ov {\bf N}}_2,{\ov {\bf n}}_0^i) + ({\ov {\bf N}}_3 ,{\ov {\bf
n}}_0^i) \nonumber\\ & & + ({\ov {\bf N}}_4,{\ov {\bf n}}_0^i)+\
({ {\bf N}}_2 ,{ {\bf 2m}}_0^i)+({ {\bf N}}_3 ,{ {\bf 2m}}_0^i)+
 ({ {\bf N}}_4 , { {\bf 2m}}_0^i)
\nonumber \eeqa

For the fixed points on the set $L_1$, the effective shift is now
$(V_9+W_9)\times (0^{m_1^i};1^{n_1^i})$. The massless fermions fall
into the following representations:
\beqa
{\bf 5_{L_1} 9} \quad {\rm Fermions_+} :\nonumber\\ & &
({\bf 2N}_0, {\bf n}_1^i) + ({\ov {\bf N}}_1,{\ov {\bf n}}_1^i )+
({\ov {\bf N}}_2,{\ov {\bf n}}_1^i) + ({\bf N}_3,{\ov {\bf
n}}_1^i) +(\ov {{\bf N}}_4,{\bf n}_1^i) \nonumber\\ & & + ({\bf N}_4,{\bf
n}_1^i) \ +
 ({ {\bf N}}_1 ,{ {\bf 2m}}_1^i)+({ {\bf N}}_2 ,{ {\bf
2m}}_1^i)+
 ({\ov {\bf N}}_3 , { {\bf 2m}}_1^i)\nonumber
\eeqa

Similarly for the $L_2$ set, the effective shift is $(V_9-W_9)\times
(0^{m_2^i};1^{n_2^i})$, and the massless fermions are:

\beqa
{\bf 5_{L_2} 9} \quad {\rm Fermions_+ :}  \nonumber\\
& &
({\bf 2N}_0,{\bf n}_2^i)
+ ({ {\bf N}}_1,{\ov {\bf n}}_2^i )+ ({\ov {\bf N}}_2,{\ov {\bf
n}}_2^i) + ({\bf N}_3,{ {\bf n}}_2^i) +(\ov {{\bf N}}_3,{\bf
n}_2^i)\nonumber\\ & & +  ({\bf N}_4,{\ov {\bf n}}_2^i)+  ({\ov
{\bf N}}_1 ,{ {\bf 2m}}_2^i)+({ {\bf N}}_2 ,{ {\bf 2m}}_2^i)+
 ({\ov {\bf N}}_4 , { {\bf 2m}}_2^i)\nonumber
\eeqa

To finish the supersymmetric part of the spectrum,  we should
write down the {\bf 55} sectors. As explained in the previous
section we use here the shift vectors defined from the $\gamma_5$
matrices and obtain:

\beqa
 {\bf 55}_{L_a} \quad {\rm Fermions_+:} \nonumber\\
& & 3\, ( 2{\bf m},\bf{ n})_a^i +  2\ ({\bf 1}, \ov
{\Yasymm }\,)_a^i\ +\  ({\bf{1}},\ov {\Ysymm}\,)_a^i
\qquad\qquad\qquad \qquad\nonumber \eeqa

In an analogous way we can also compute the nonsupersymmetric
massless spectrum for the anti-brane sectors. For the fermionic
part we have to keep in mind that the chirality is opposite from
that of the supersymmetric sector, and then the spectrum comes
from considering the effective shifts with $q_a^j$ and $p_a^j$
replacing  $n_a^i$ and $m_a^i$ respectively.  For instance, for
the ${\ov 5 9}$ sector, the effective shift is $V_9\times
(0^{p_0^i};1^{q_0^i})$, and so on. We find:
\beqa
 {\bf {\ov 5}_{L_0}9} \quad {\rm Fermions_-} :\nonumber \\
& & ({\bf 2N}_0,{\bf q}_0^j) + ({\ov {\bf N}}_1,{\bf q}_0^j)+
 ( {\bf N}_1,{\bf q}_0^j ) +
({\ov {\bf N}}_2,{\ov {\bf q}}_0^j) + ({\ov {\bf N}}_3 ,{\ov {\bf
q}}_0^j) \nonumber \\ & &+({\ov {\bf N}}_4,{\ov {\bf q}}_0^j)+ ({
{\bf N}}_2 ,{ {\bf 2p}}_0^j)+({ {\bf N}}_3 ,{ {\bf 2p}}_0^j)+  ({
{\bf N}}_4 , { {\bf 2p}}_0^j) \nonumber\\ {\rm Scalars }
:\nonumber \\ & & ({\bf 2N}_0,{\bf 2p}_0^j) + ({\ov {\bf
N}}_1,{\bf 2p}_0^j)+({{\bf N}}_1,{\bf 2p}_0^j)+ \nonumber\\ & &
[({\ov {\bf N}}_2,{ {\bf q}}_0^j) + ({\ov {\bf N}}_3 ,{{\bf
q}}_0^j)+ ({\ov {\bf N}}_4,{{\bf q}}_0^j) +\quad h.c] \nonumber
\eeqa
\beqa
{\bf {\ov 5}_{L_1} 9} \quad {\rm Fermions_-} :\nonumber\\ &
& ({\bf 2N}_0, {\bf q}_1^j) + ({\ov {\bf N}}_1,{\ov {\bf q}}_1^j )
+ ({\ov {\bf N}}_2,{\ov {\bf q}}_1^j)+ ({\bf N}_3,{\ov {\bf
q}}_1^j) +(\ov {{\bf N}}_4,{\bf q}_1^j) \nonumber \\ & & +({\bf
N}_4,{\bf q}_1^j)+ ({ {\bf N}}_1 ,{ {\bf 2p}}_1^j)+({ {\bf N}}_2
,{ {\bf2p}}_1^j)+
 ({\ov {\bf N}}_3 , { {\bf 2p}}_1^j) \nonumber\\
 {\rm Scalars} :\nonumber\\ & &
({\bf 2N}_0, {\bf 2p}_1^j) +(\ov {{\bf N}}_4,{\bf 2p}_1^j)+ ({\bf
N}_4,{\bf 2p}_1^j) \nonumber \\ & & + [({\ov {\bf N}}_1,{ {\bf
q}}_1^j )+ ({\ov {\bf N}}_2,{{\bf q}}_1^j) +({\ov {\bf N}}_3,{{\bf
q}}_1^j)+ \quad h.c.]\nonumber \eeqa
\beqa
{\bf {\ov 5}_{L_2} 9} \quad {\rm Fermions_-} :\nonumber\\ &
& ({\bf 2N}_0,{\bf q}_2^j) + ({ {\bf N}}_1,{\ov {\bf q}}_2^j ) +
({\ov {\bf N}}_2,{\ov {\bf q}}_2^j) + ({\bf N}_3,{ {\bf q}}_2^j) +
(\ov {{\bf N}}_3,{\bf q}_2^j) \nonumber \\ & & + ({\bf N}_4,{\ov
{\bf q}}_2^j) + ({\ov {\bf N}}_1 ,{ {\bf 2p}}_2^j)+({ {\bf N}}_2
,{ {\bf 2p}}_2^j)+ ({\ov {\bf N}}_4 , { {\bf 2p}}_2^j) \nonumber\\
{\rm Scalars} :\nonumber \\ & & ({\bf 2N}_0,{\bf 2p}_2^j)+ ({\bf
N}_3,{ {\bf 2p}}_2^j) + (\ov {{\bf N}}_3,{\bf 2p}_2^j) \nonumber
\\ & & + [({ {\bf N}}_1,{{\bf q}}_2^j )+({ {\bf N}}_2,{\ov {\bf
q}}_2^j) + ({\bf N}_4,{ {\bf q}}_2^j)+ \quad h.c]\nonumber \eeqa
\newpage
\beqa
 {\bf {\ov 5}{\ov 5}}_{L_a} \quad
{\rm Fermions_-} :\nonumber\\
& & \ (\, {\Yasymm};1)_a^j \ +\ (\, 1;Adj)_a^j\ \nonumber \\
 {\rm Fermions_+}:\nonumber\\
& & 3(2{\bf p} , \bf{q })_a^j + 2\ (\ 1,\ov {\Ysymm }\,
)_a^j\  +\ (\, 1,\ov {\Yasymm}\,)_a^j\nonumber \\ \quad {\rm
Scalars:} \nonumber\\ & & 3\, ( 2{\bf p},\bf{ q}\ )_a^j +  2\
({\bf 1}, \ov {\Yasymm }\,)_a^j\ +\  ({\bf{1}},\ov {\Ysymm}\,)_a^j
\qquad\qquad\qquad\nonumber
\eeqa

As we  mentioned, there can also exist  branes (or antibranes) not attached to fixed points.
For instance, for $r $ coincident antibranes travelling in the bulk
a $Sp(2r)$ gauge group appears.  The ${\bf {\ov 5}{\ov 5}}$ massless
spectrum is
\beqa
{\bf {\ov 5}{\ov 5}}_{bulk} \quad
{\rm Fermions_-} :\nonumber\\
& & \ ( {\Yasymm}\,)\ \nonumber \\
{\rm Fermions_+}:\nonumber \\
& & 2\ ({\bf (2r+1)r})+ \ ({\Yasymm }\,)
 \nonumber \\
{\rm Scalars:} \nonumber \\
 & &  2\ ({\Yasymm }\,) + ({\bf (2r+1)r})
 \qquad\qquad\qquad\ \qquad\qquad\qquad \nonumber
\eeqa

The ${\bf {\ov 5} 9}$ sector contains negative chirality fermions and bosons in
$$
({\bf 2N}_0,{\bf 2r}) + [({\bf N}_1,{\bf  2r})+({{\bf N}}_2,{{\bf
2r}})
 +({{\bf N}}_3,{\bf 2r})+({{\bf N}}_4,{{\bf 2r}})+ {\bf N}_j
 \rightarrow {\ov {\bf N}}_j\,]$$
multiplets.\footnote{In the presence of certain discrete Wilson lines there
are no massless particles in the ${\bf {\ov 5} 9}$ sectors and
hence the ${\bf {\ov 5}{\ov 5}}$ states  act as sort of
hidden sectors for the ${\bf 99 }$ particles. See the discussion at the
beginning of section 4.1.} Recall that the $r(2r-1)$ dimensional  $Sp(2r)$ antisymmetric
representation is actually reducible to a $r(2r-1)-1$ dimensional
 representation plus a singlet.

In the spectrum above we have not indicated the corresponding
$U(1)$ charges. The convention is that a fundamental of $U(N)$,
$\bf N$ carries charge 1 (no normalized ) with respect to $U(1)$
in $U(N)$ while ${\bf \ov N}$ has charge $-1$. Thus, for instance,
$({\ov {\bf N}}_3,{\bf n}_2^i)$ carries charge $-1$ with respect
to $U(1)$ in $U(N_3)$, and $1$ with respect to the $U(1)$ groups
in $U(n_2^i)$.
 Also an antisymmetric $ {\Yasymm }$ representation
carries charge $2$ while $\ov {\Yasymm }$ has charge $-2$.

It is straightforward to check that all non-abelian gauge
anomalies cancel if tadpole conditions (\ref{tad}) are satisfied.
For this we have to keep   in mind that branes and anti branes do
not coexist at a given point so, for instance, if for a given $i$,
we have $n_a^i\neq 0$ this implies that $q_a^i=0$. Also see that
multiplicities coming from ${\bf {9}{\ov 5}}_{L_a}$ sectors appear
with a negative sign since fermions carry opposite chirality.

{}From this general expressions we can consider special cases which
simplify substantially the spectrum. For instance in some models
the tadpoles cancel for one or more of the sets $L_a$, so there is
no need of introducing 5-branes and antibranes at those points. In
those cases the coefficients $m_a^i,n_a^i,p_a^j,q_a^j$ will vanish
and also their corresponding gauge groups and matter content. A
particularly interesting case is when we set all the antibranes in
the bulk, so that only 5-branes are trapped at  some of the fixed
points ($r=2\sum_{i,a}(n_a^i+m_a^i)$). We will see explicit
examples of these cases in the following sections.

\subsection{Anomalous $U(1)$ cancellation and FI-terms}

Knowing the general spectrum we can see that there are several
$U(1)$ symmetries. As usual in string theory some of these
symmetries are anomalous but the anomaly is cancelled by the
standard Green-Schwarz mechanism. Unlike the heterotic models,
 type I models have  the peculiarity of
having several twisted  Ramond-Ramond fields which can participate
in the anomaly cancelling mechanism, therefore allowing for the
existence of several anomalous $U(1)$'s. In the heterotic case
only one $U(1)$ symmetry could be anomalous since there is only
one antisymmetric tensor field to cancel it. For example in a
model with only 9-branes the  coupling of the  $U(1)$ fields to
the antisymmetric Ramond-Ramond field in the $k$-th twisted sector
$B_{k}^{\mu\nu}$ is proportional  to \cite{iru}:
\beq
i\Tr (\gamma_k\lambda_i) B_k\wedge F_{U(1)}\ =\ (2n_i \sin
2\pi k V_i) \  B_k\wedge F_{U(1)} \eeq
where $n_i$ is the  rank of the $ith$ U(n) group and $V_i$ is the
component of the corresponding shift vector overlapping with that
$U(n)$. Furthermore, supersymmetry  implies that the anomaly
cancellation term induces a Fayet-Iliopoulos term of the form:
\beq
-D_i\ \sum_f\ n_i \sin 2\pi k V_i \ M_f \eeq
where the sum runs over the fixed points.
 $M_f$ is the Neveu-Schwarz partner of the Ramond-Ramond
 antisymmetric tensor field (which together form a supersymmetric
 linear multiplet in four dimensions, usually dualized to a standard
 chiral multiplet).
The net effect of the anomaly cancelling and Fayet Iliopoulos
terms is to give a mass \cite{iru,pop,lln, celw}
to the anomalous gauge field and the
multiplet
 containing $M_f$ and $B_f^{\mu\nu}$.
 Contrary to heterotic models, in  type I models
the anomalous $U(1)$'s can survive at low energies as global
symmetries, which could have interesting phenomenologically
 implications
\cite{iq}.

Let us then identify which of the many $U(1)$ symmetries of our
models are anomalous. For definiteness we will concentrate
explicitly on the models with only trapped branes (anti-branes on
the bulk). First of all we notice that each of the $U(1)$'s is
inside a $U(n)$ symmetry. There are four of them coming from the
9-brane sector groups $U(N_s)$, $s=1,\cdots , 4$ and a maximum of
 9 coming from the 5-branes groups $U(n_a^i)$. We need to compute the
mixed  $U(1)-SU(N_s)^2$, $U(1)-SO(2N_0)^2$,  $U(1)-SU(n_a^i)^2$
and $U(1)-Sp(m_a^i)^2$ anomalies, which is
then  a  matrix with 13 rows corresponding to the $4+9$ $U(1)$'s
and 23 columns corresponding to the nonabelian groups
$SU(N_s)$, $SO(2N_0)$, $SU(n_a^i)$ and $Sp(2m_a^i)$. It  can be labeled as
$T^{\alpha\beta}_{IJ}$ where the super-indices $\alpha,\beta=9,5$
only label the brane origin of the groups. The matrix indices
$I,J$ label the anomaly of the $Ith$ $U(1)$ group
with the $Jth$ nonabelian group $G_J$.
 Looking at the spectrum we can simplify the
writing of the  matrix since for a given fixed point sector $L_a$ the
corresponding $U(n_a^i)$ and $Sp(m_a^i), i=0,1,2$, are essentially the
 same.
 The
anomaly matrix is then an  array with the first 4 rows giving the
 anomalies
for the $U(1)$'s of the 9-brane sector and the last three for the
$n_0^i, n_1^i, n_2^i$. The 11 columns would then correspond to: the
 first 4 to $SU(N_s)$, the 5th to $SO(2N_0)$, th next three to
 $SU(n_a^i),
a=0,1,2$ and the last three to $Sp(2m_a^i), a=0,1,2$.

{}From the spectrum above we find:

 \begin{equation}
T_{IJ}^{\alpha\beta} =  \frac{1}{2}\left ( \begin{array}{ccccccccccc} 0 & 0 &
-3N_1 & 3N_1 & 0  & 0 & -N_1 & N_1 & 0 & N_1 & -N_1  \\
 0 & -3N_2 & 0 & 0 & 3N_2 & -N_2 & -N_2
& -N_2 & N_2 & N_2 & N_2   \\
 3N_3 & 0 & 0 & -3N_3 & 0 & -N_3 & N_3 & 0 & N_3 & -N_3 & 0   \\
 3N_4 & 0 &
-3N_4 & 0 & 0 & -N_4 & 0 & N_4 & N_4 & 0 & -N_4   \\
2n_0^i & -n_0^i & -n_0^i & -n_0^i
& n_0^i & -3n_0^i & 0 & 0 & 3n_0^i & 0 & 0 \\
 -n_1^i & -n_1^i & -n_1^i & 2n_1^i & n_1^i & 0 &
-3n_1^i & 0 & 0 & 3n_1^i & 0  \\
 -n_2^i & -n_2^i & 2n_2^i & -n_2^i & n_2^i &  0 & 0 &
-3n_2^i & 0 & 0 & 3n_2^i \\
\end{array}
\right )
\end{equation}
where  the  dependence on  the ranks of the different $U(n)$
groups reflects the fact that the $U(1)$'s are not normalized. One
can check that in the more general case there are up to 9
anomalous $U(1)$'s  and 4 non-anomalous $U(1)$'s. The 4 anomaly
free combinations have the following general form:
\beqa
b_1{{Q_{N_1}}\over {N_1}} & + & b_2{{Q_{N_2}}\over {N_2}}\
 +\ b_3{{Q_{N_3}}\over
{N_3}}\ +\ b_4{{Q_{N_4}}\over {N_4} }\ +  \\
 -{1\over 3} (b_2+b_3+b_4) \sum _a {{Q_{n_a^0}\over {n_a^0}}}\ & +&
{1\over 3} (-b_1-b_2+b_3) \sum _a {{Q_{n_a^1}\over {n_a^1}}}  +
{1\over 3} (b_1-b_2+b_4) \sum _a {{Q_{n_a^2}\over
{n_a^2}}}\nonumber \label{sinanom} \eeqa
where the $b_i$ are arbitrary real coefficients.
A linearly independent
 set of the four anomaly free $U(1)$'s can  be written as:
\beqa
Q_A & = & Q_{N_4} - Q_{N_3}- Q_{N_1}\nonumber \\ Q_B & = &
\sum_{a,i=0}^2 Q_{n_a^i} - 3Q_{N_2}\nonumber\\ Q_C & = & 3\
\sum_{i=0}^2 Q_{n_0^i} - \sum_{a,i=0}^2\  Q_{n_a^i}  - 3 Q_{N_3}-
3Q_{N_4}\nonumber\\ Q_D & = & \sum_{i=0}^2\ \left( Q_{n_1^i} -
Q_{n_2^i}\right)\ - 2Q_{N_1} + Q_{N_3} - Q_{N_4}
\label{sinanomlin}
 \eeqa
where, to simplify the expressions,  we have absorbed
 the rank in the definition of the charges ($Q_{N_1}/N_1\rightarrow
 Q_{N_1}$
and so on) \footnote{In some particular cases several of the
 abelian or non-abelian symmetries are not present and the
 corresponding modifications of the anomaly free and anomalous
 symmetries have to be done.}. 
It is straightforward, though tedious, to verify that 
all cubic $U(1)$ anomalies cancel for these combinations. 
  $Q_A$ comes completely from the 9-brane sector and it
will  be eventually identified
 with hypercharge in the Standard-like models of
the  next sections.
The  9 anomalous $U(1)$'s are then:
\beqa
Q_{N_2}+Q_{N_3}+Q_{N_4}&+&\sum_{i=0}^2 Q_{n_0^i},  \\
\nonumber Q_{N_1}+Q_{N_2}+Q_{N_3}&+&\sum_{i=0}^2 Q_{n_1^i}, \\
\nonumber Q_{N_2}-Q_{N_1}+Q_{N_4} &+&\sum_{i=0}^2
Q_{n_2^i},\nonumber\\ Q_{n_a^1}-Q_{n_a^2}, & &
Q_{n_a^1}+Q_{n_a^2}-2Q_{n_a^3} \qquad a=0,1,2 \eeqa
The presence of at most 9 anomalous $U(1)$'s is expected because
there are only 9 fixed points in the first two complex planes, and
hence there are 9 RR twisted fields which can participate in the
generalized Green-Schwarz mechanism. The anomalous $U(1)$'s will
become massive by using the 9 RR twisted field combinations as
their longitudinal components so that they become massive vector
supermultiplets. As discussed above, Fayet-Iliopoulos terms
depending on the twisted NS fields do also appear. Let us write
the above 9 anomalous $U(1)$ generators in the general formula
\beq
Q^r\ =\ \sum _{s =1}^4 \ d_{s }^r Q_{N_{s}}\ +\ \sum
_{a,i=0}^2\ f_{a,i}^r Q_{n^i_a} \label{genanom} \eeq
where $r$ goes from 0 to 8 and the coefficients $d_{s }^r$
vanish for $r>2$. Then each of the $U(1)^r$ will have a
Fayet-Iliopoulos term  $\xi ^r$ given by \cite{iru,iru2}
\beqa
\xi ^r \ &  = & \ \sum _{s =1}^4\ \sum_{\alpha = 1}^{N_s} \ d_{s}^r
\sum_{a,i=0}^2\sin[2\pi (V_9+aW_9)_{\alpha}] \ M_{(a,i)} \nonumber \\
& &  +\ 4\sin(\pi v_1)\sin(\pi v_2)
\sum _{a,i=0}^2 \ f_{a,i}^r \sin[2\pi V_5]_{(a,i)} \ M_{(a,i)}
\label{fi9} \eeqa
where the $M_{(a,i)}$, $a,i=0,1,2$ are the twisted NS fields
associated to the first two complex planes. The above are nine
linear independent combinations of the nine twisted moduli.
A nontrivial check of the validity of this expression is that
it gives $\xi=0$ for all the anomaly free combinations, as it should.
 In the
effective scalar potential there will be terms coming from the
D-terms of the anomalous $U(1)$'s which will have the general
qualitative structure
\beq
V_r\ =\ {1\over 2} (\sum _l q^r_l|\phi _l|^2\ +\ \xi ^r)^2
\label{potfi}
 \eeq
where $\phi _l$ denote scalar fields with charge $q_l^r$ under
$U(1)^r$. Notice that putting all charged fields to zero and also
the  twisted modes
 $(\xi ^r=0)$ is  perfectly consistent. Thus unlike
what happens in perturbative heterotic vacua \cite{dsw}
 the FI-terms do not
necessarily trigger further gauge symmetry breaking and the anomalous
$U(1)$'s may remain as global symmetries at low energies. On the other
hand one can also obtain $V_r=0$ for configurations with some
non-vanishing vev for some $\phi _l$ fields by compensating it by
appropriately choosing the $\xi ^r$ values. Thus in the specific
examples of section 4 we will see how in the $(55)$ sectors there
are singlet scalars charged under the anomalous $U(1)$'s. They can
be given non-vanishing vevs as long as we also give appropriate
vevs to some particular combinations of twisted NS fields.

Let us come back now to the issue of $U(1)$ anomaly cancellation.
The anomaly cancelling coefficients have been worked out for the
general class of $\IZ_N$ orientifold models and found to take the
form \cite{iru}.
\beq
A^{\alpha\beta}_{IJ}\ =\  \ {1\over N}\ \ \sum_{k=1}^{N-1} \
C_k^{\alpha\beta}(v)\ n_I^\alpha\sin2\pi kV^{\alpha}_I \ \cos2\pi
kV^{\alpha}_J \label{masterorient} \eeq

Here $k$ runs over twisted $\IZ_N$ sectors, $\alpha$, $\beta$ run
over 5,9, (meaning  5- or 9-brane origin of the gauge boson), 
$V^{\alpha}_I$ is the component of the corresponding shift
vector $V_9, V_9+W_9$ or $V_9-W_9$, along the entries overlapping
with the corresponding  group. Notice that
since Wilson lines are included, a sum over
contributions from different fixed points is understood in the
expression above, each set of fixed points $L_a$ 
feeling a different shift vector. Finally 
\beqa
C_k^{\alpha\alpha} & = & \prod _{a=1}^3 2\sin\pi kv_a \quad
{\rm for}\;\; \alpha=\beta \nonumber\\ C_k^{59 } & = & 2\sin\pi
kv_3 \label{ckpp59}
\eeqa
Here $v_a=\ell_a/N, a=1,2,3$
 specify the orbifold twist on each of the complex planes, as
introduced in the previous section. 
The same expressions are valid for antibrane-antibrane
sectors but an extra minus sign must be included because of
opposite chirality. 
For the $\IZ_3$ case under consideration we have
 $C_1^{\alpha\alpha}=-C_2^{\alpha\alpha}=-3\sqrt{3}$ and
$C_1^{59}=-C_2^{59}=-\sqrt{3}$, therefore the contributions of the
two
 sectors $k=1,2$ are identical and we find:
\beq
A^{\alpha\beta}_{IJ}\ =\ -T^{\alpha\beta}_{IJ}\
\eeq
therefore anomalies are exactly cancelled as expected. Similarly
we can compute the mixed $U(1)$ anomalies which are also cancelled
by the Green-Schwarz mechanism. Mixed $U(1)$-gravitational
anomalies can also be computed. We obtain from the triangle diagrams:
$ (0,-9N_2,0,0,-3n_0^i,-3n_1^i,-3n_2^i)$  which
is exactly cancelled 
 by the general coefficient \cite{iru}:
\beq
A_I^\alpha=\frac{3}{4N}\, \sum_\beta\, \sum_k\
C_k^{\alpha\beta}(v) n_I\sin 2\pi k V_I^\alpha \Tr\left(
(\gamma_k^\beta)^{-1}\right)
\eeq
where again the $V_I$ is the relevant component of each of the three
different shifts. This cancellation 
 can be considered as a nontrivial consistency check for the
whole construction.

\section{Examples of three-generation models}

As we discussed in the previous section, untwisted tadpole
conditions require the same number of $5$-branes and
anti-$5$-branes in the vacuum. We will consider two general type
of models in turn: i) Models with only 5-branes stuck at fixed
points and ii) Models with both $5$-branes and anti-$5$-branes
present at fixed points
\footnote{There is a third logical possibility in which 
only anti-$5$-branes are stuck at the fixed points and
$5$-branes live in the bulk. In this case one obtains 
non-supersymmetric models which are very similar in their characteristics to
the type ii) above. Thus we will not consider this option in the
remainder of this paper.}. The first class of models lead
generically to ``gravity'' mediation of SUSY-breaking. In this
case the natural value for the string scale is the intermediate
scale. The second class of  models are  explicitly
non-supersymmetric.
 This requires lowering the
string scale further, as we discuss in section 5. We will describe
these two types in turn.

Using the general formalism of section 3, it is straightforward to
look for values for $N_s$ and $n_a^i, m_a^i$ that give
standard-like models. The groups will generically be much larger
than $SU(3)\times SU(2) \times U(1)$, but we can look for models
for which some of the gauge sectors are hidden, in the sense that
they do not have direct non-abelian couplings with the standard
model fields.
 It turns out to be difficult to accommodate the group
of the SM and the necessary left-handed quarks inside a $(55)$
sector and hence we will embed
  the full standard model group
inside  the 9-branes sector. We would not get the right spectrum
from $SO(2N_0)\times SU(N_2)$ part neither, therefore we are
limited to a subgroup of $U(N_1)\times U(N_3)\times U(N_4)$.
Furthermore, it is easy to see that the spectrum is invariant
under permutations of $N_1,N_3$ and $N_4$, so without loss of
generality we can identify say $N_3=3, N_4=2, N_1=1$ to give us
the $SU(3)\times SU(2)\times U(1)$ respectively (we may have to
redefine fundamental and anti-fundamental representations in some
cases). Plugging these values into the tadpole cancellation
conditions we find that $N_2$ and $n_a^i$, $m_a^i$, $p_a^j, q_a^j$
have to satisfy:
\beqa
7-N_2 & = & n_0^i-2m_0^i \, = \ 2p_0^j-q_0^j \nonumber\\
8-N_2 & = & n_1^i - 2m_1^i\, =\, 2p_1^j - q_1^j \nonumber \\ 9-N_2
& = & n_2^i - 2 m_2^i\, = \, 2p_2^j- q_2^j \eeqa
Therefore the configurations with the minimum number of 5-branes
\footnote{As argued in chapter 6, one may expect that  configurations 
with small  number of brane-antibrane pairs in the vacuum to be 
dynamically preferred with respect to those with large numbers
since they contribute to the vacuum energy in a way proportional to their
tension.} 
correspond to

\begin{enumerate}

\item  $N_2=8$, $N_0=2$. The 9-brane group is $U(3)\times U(2)\times
U(1)\times U(8)\times SO(4)$. The number of 5-branes or anti
5-branes can be such that $n_0^i=m_0^i=n_2^i=1$, similar to the
previous case or $n_2^i=q_0^i=1$ with two 5-branes at each of the
points $L_2$ and two anti 5-branes at each of the points $L_0$.

\item
 $N_2=7$, $N_0=3$. The 9-brane group is $U(3)\times U(2)\times
U(1)\times U(7)\times SO(6)$. We may have four 5-branes at each of
the points $L_2$ ($n_2^i=2$) or two anti 5-branes ($p_2^j=1$) and
two  5-branes at each of the points $L_1$ ($n_1^i=1$).

\item  $N_2=9$, $N_0=1$ for which the 9-brane symmetries are
$U(3)\times U(2)\times U(1)\times U(9)\times U(1)$. The
arrangement of 5-branes and anti 5-branes can be done in at least
two minimal ways: $n_1^i=m_1^i=m_0^i=1$, all other $n_a^i, m_a^i,
p_a^j, q_a^j$ vanishing. This corresponds to two 5-branes at three
fixed points $(L_0)$ and four 5-branes at other three fixed points
$(L_1)$. Alternatively we can set two 5-branes at the points
$L_0$($m_0^i=1$)
and two anti 5-branes at each of the points in $L_1$ ($q_1^i=1$).
(we could also choose a more complicated configuration such as
$q_0^j=2, n_1^i=m_1^i=1$ corresponding to
four anti 5-branes at three fixed points ($L_0$) and four 5-branes
at each of the other fixed points ($L_1$)).

\end{enumerate}

In each of the three cases the spectrum and 5-branes gauge groups
are different implying different phenomenology. We will discuss
below the first case ($N_2=8$) in detail  and leave the other two
for the appendix. The three of them are summarized in the tables below.

Another important issue is the way hypercharge is embedded into
the gauge group of the models. Hypercharge will be identified with
one of the anomaly-free $U(1)$'s in eq.(\ref{sinanom}). The
simplest such embedding correspond to
$(b_1,b_2,b_3,b_4) =(1,0,1,-1)$ which indeed gives the
appropriate hypercharge of the left-handed quarks which reside in
the $(99)$ sectors of these models. In this case one sees from
eqs.(\ref{sinanom}) that the hypercharge generator is purely
embedded into the $(99)$ sector without any mixing from $(55)$
sectors.

One can in principle try to identify the hypercharge generator
with other combinations which also give the correct hypercharge of
left-handed quarks, like $(b_1,b_2,b_3,b_4) =(-1,0,-2,1)$. In
this case eqs.(\ref{sinanom}) show that the hypercharge generator
will mix with the $U(1)$'s coming from $5$-branes. This means in
particular that the fields in the $(55)$ sector will have
(fractional) electric charges. The same will be true in models
with $N_1=0$. In those models the $(99)$ sector will only contain
the left-handed quarks and the $U(1)$ generator giving correctly
its hypercharge will necessarily mix with $U(1)$'s coming from
$5$-branes. In fact,  fields coming from the $(55)$ sectors will
in general get vacuum expectation values as soon as the blowing-up
modes are not strictly zero
(see eq.(\ref{potfi})). Then only the hypercharge fully
embedded in the $(99)$ sector remains unbroken and the other
possibilities are not really viable since the corresponding
generators are broken. Thus we will identify the physical
hypercharge generator with $Q_A$ in (\ref{sinanomlin}).

A similar analysis can be done if we want a left-right symmetric
model $SU(3)_c\times SU(2)_L\times SU(2)_R$, for which we want
$N_3=3, N_1=N_4=2$. In this case, there are two simple options:
$N_2=7$, $N_0=2$ which in order to satisfy the tadpole equations
require at least $n_2^i=1$ which means two 5-branes trapped at
each of the three fixed points in $L_2$. This is probably the
simplest model and we will discuss it in detail next. The second
option requires $N_2=8$, $N_0=1$ which is a bit more complicated
since it requires at least $n_0^i=m_0^i=n_1^i=m_1^i=1$, which means
four 5-branes at each of the six points in $L_0,L_1$, or substituting
them by anti 5-branes in the three points in $L_1$ and one of
the points in $L_0$. We will not discuss this model in detail, but only
present its spectrum in the tables.

If we want to have larger symmetry groups, we may consider a
$SU(4)\times SU(2)_L\times SU(2)_R$  model
of the Pati-Salam type. For this we then choose $N_3=4$,
$N_1=N_4=2$. Again there are two simple solutions. One with
$N_2=6$, $N_0=2$ requires only $n_2^i=2$ which is four 5-branes at
each of the three fixed points $L_2$. We will discuss this model
and its realization in terms of trapped branes and anti branes in
the next subsection. The second option $N_2=8$, $N_0=0$ is also
very simple, requiring only $m_1^i=m_0^i=1$ which is two 5-branes
at each of the six fixed points $L_0$ and $L_1$. The spectrum can be
seen in the tables below. We will write the
full massless spectrum in the tables. Symmetries even larger
than Pati-Salam, like a standard GUT group (see the example  in the
appendices)  or flipped $SU(5)$ do
not seem to have interesting matter content in this class of models.

We will discuss next in some detail several standard-like models,
first the class with only branes trapped at the fixed points
(gravity mediated)
and then the one with both branes and anti-branes trapped at the fixed points
with explicit supersymmetry breaking. We summarize the spectrum of the
three
$SU(3)\times SU(2)\times U(1)$ models, the two $SU(3)\times
SU(2)_L\times SU(2)_R$ models and the two Pati-Salam models
mentioned above in
two tables, one for each class.

\subsection{ Models with only branes stuck at fixed points}

In order to cancel twisted tadpoles it is enough to  add
$5$-branes appropriately to the fixed points in the first two
complex planes of the underlying $\IZ_3$ orbifold. The
anti-$5$-branes will be located generically in the bulk of the
first two complex planes. In addition one can always add an
appropriate Wilson line in the third complex plane and acting only
on anti-$5$-branes in such a way that they are decoupled both from
$9$-branes and $5$-branes \footnote{This procedure is more
transparent if we perform a T-duality transformation in the third
complex plane in such a way that $9$-branes turn into $7$-branes
localized at the origin in the third complex plane. The
anti-$5$-branes turn into  anti-$3$-branes. The latter decouple
from $7$-branes if we locate them away from the origin.}. In this
way the anti-$5$-brane sector (which is the only one which is not
supersymmetric) behaves like a sort of "hidden sector" for the
"observable" sector of $9,5$-branes which is supersymmetric.
Supersymmetry-breaking effects in this visible sector will be
suppressed by the compactification radii providing an example of
gravity mediated SUSY-breaking. As we already mentioned one could
worry about  the presence of attractive forces between branes and
antibranes which might render this type of configuration unstable.
We postpone the discussion of these effects to chapter 5 and we
limit ourselves to the construction of these models (otherwise
tachyon-free and free of RR tadpoles).

\bigskip

{\bf i) A three generation standard model}

\bigskip

Let us consider the $\IZ_3$ orientifold with an action of the twist
on the Chan-Paton factors given by $N_0=2, N_1=1, N_2=8, N_3=3,
N_4=2$ as  mentioned above.
 We choose to reorder the twist and
Wilson lines in order that the Standard Model sector can be
clearly distinguished, so we will write:
\beq
V_9 \ = \
1/3(1,1,1,-1,-1,0,0,0,1,1,1,1,1,1,1,1) \eeq A quantized Wilson
line is added in the first complex plane given by:
\beq
W_9 \ = \
1/3(1,1,1,1,1,-1,0,0,0,0,0,0,0,0,0,0) \eeq The gauge group from
the (99) sector will be $U(3)\times U(2)\times U(1)\times
[SO(4)\times U(8)]$. Now, the 9  $\IZ_3$ fixed points in the first
two complex planes split into three sets of three fixed points
each : $(0,m)$, $(1,m)$ and $(-1,m)$ with $m=0,\pm 1$. The shifts
corresponding to each of these three sets are $V$, $V+W$ and $V-W$
respectively. The corresponding value for $\Tr \gamma_{\theta,9}$
are -7, -4 and -1 respectively. Looking at the tadpole
cancellation conditions we see that we will be forced to add
$5$-branes at the fixed points $(0,m)$ with $\Tr
\gamma_{\theta,5}= +1$ and also at the fixed points $(-1,m)$ with
$\Tr \gamma_{\theta,5}= -1$. No $5$-branes have to be added to the
points $(1,m)$ since tadpoles cancel without them. Thus the
simplest option is to locate 5-branes with twist actions given by:
\beqa
(0,m)   \rightarrow  \  \gamma _{\theta,5}  & = & \diag  (
I_{2},\alpha ,\alpha^2 ) \nonumber\\ (-1,m)   \rightarrow \  \gamma
_{\theta,5}  & = & \diag  (\alpha ,\alpha^2 ) \eeqa Notice that
these $5$-branes are stuck at the fixed points, they cannot leave
the fixed points without violating the twisted tadpole conditions.
Let us then compute the massless spectrum. After an analysis of
the $U(1)$'s in the model one can check that the generator given
by \footnote{Notice that this definition corresponds to the $Q_A$
generator of eq.(\ref{sinanomlin}) once one takes into account the
flip in sign in the twist vectors $V$ and $W$ for the
$U(2)$ and $U(1)$ groups compared to the conventions of chapter 3.}
:
\beq
Y\ =\ (-1/3,-1/3.-1/3,-1/2,-1/2,1,0,..,0)
 \eeq
is anomaly
free and can be identified with standard weak hypercharge. One
finds the following chiral charged fields from each sector:

{\bf (99):}

There are  chiral multiplets  transforming under the gauge group
like \footnote{Properly speaking, using the general formulae
of chapter 3 we would have obtained the conjugate states to the
ones we display in this chapter 4. We have preferred to list
the conjugates so that one recovers the usual conventions
of the SM in which it is left-handed quarks which are
colour triplets (and not antitriplets).}  :
\beq
 3(3,2,1/6)\
+\ 3({\bar 3},1,-2/3) \ +\ 3(1,2,+1/2) \ +\ [ 3(1,28)' +3(4,{\bar
8})'] \eeq These fields  correspond to 3 generations of left-handed
quarks and right-handed U-quarks as well as three sets of Higgs
fields $H_U$.

{\bf (59) :}

{}From the 5-branes at the $(-1,m)$ fixed points we get a $U(1)_5^3$ gauge
group. One then finds  chiral multiplets (in three copies, one per
fixed point) transforming like:
\beq
(1,2,-1/2)_q \ +\ (1,1,+1)_q
\ +\ (3,1,-1/3)_{-q}+({\bar 3},1,1/3)_{-q} \ + \ [ (1,8)'_{q}\ +\
(4,1)'_{-q}] \eeq where the subindex $q$ refers to the $U(1)^3_5$
charges. These fields include three standard lepton generations
and three sets of vector-like quarks. The $SU(2)$ doublets could
also be $H_D$ type of Higgsses which have same quantum numbers as
left-handed leptons.

{}From the 5-branes at the $(0,m)$ fixed points we get a 
$(Sp(2)\times U(1))^3$
gauge group. One then finds  chiral multiplets (in three copies,
one per fixed point) transforming like:
\beqa
 (1,2,-1/2;2)&
+&(1,2,1/2;1)_r \ + \
 ({\bar 3},1,1/3;2)\  + \
(3,1,-1/3;1)_r\ +\nonumber\\
  (1,1,+1;1)_{-r} \ & + &\  (1,1,-1;1)_{-r} \
 +  \ [ (1,{\bar 8};2)'+(1,8)'_r\ +\ (4,1)'_{-r}]
\eeqa where the subindex $r$ refers to the $U(1)^3_5$ charges. The
doublets after the semicolons refer to the $Sp(2)$ groups. These
chiral multiplets include a net number of three right-handed
D-quarks and three left-handed leptons (or $H_D$ doublets). In
addition there are three pairs of vector-like right-handed
$D$-quarks and leptons.

{\bf (55) :}

For those $5$-branes sitting at fixed points $(-1,m)$ there is
just one singlet chiral field  $(1)_{2q}$ associated to the third
complex plane. For those $5$-branes sitting at fixed points
$(0,m)$ there are multiplets transforming under $Sp(2)\times U(1)$
like:
\beq
3(2)_{-r}\ +\ (1)_{2r} \eeq

Altogether this model has three standard model quark lepton
generations and in addition 6 copies of Higgs-like fields
$(H_U+H_D)$ plus 6 vector-like sets of $D$-quarks and three
vector-like sets of right-handed electrons. Vacuum expectation
values of singlets in $(55)$ sector can give masses to many of
these extra particles. In particular, as we will mention below, in
the $(55)$ sector of $5$-branes on the fixed points $(-1,m)$ there
are singlets which couple to the extra triplets in that $(95)$
sector. If those singlets get a vev (which may happen in a D-term
flat manner if we turn on a vacuum expectation value to the
relevant blowing-up fields) those triplets get heavy.
The same happens with the extra right-handed leptons in
the $(59)$ sectors corresponding to fixed points $(0,m)$.
We would be thus left with 3 standard quark-lepton
generations plus 6 sets of extra vector-like  left-handed
leptons and 3 sets of vector-like $D_R$ quarks.
  We will see
in chapter 5 that in that case gauge coupling unification takes
place at the intermediate scale precisely due to the presence of
these extra fields.

Let us finally mention a few aspects with regard to the $U(1)$'s
in this model. Using the results of chapter 3 it is easy to check
that there are 7 anomalous $U(1)$'s (which then will become
massive as usual). There remain 3 anomaly-free $U(1)$'s, one of
which is the hypercharge defined above. The other two anomaly-free
$U(1)$'s, which involve a mixture with $5$-brane $U(1)$'s,   will
also be spontaneously broken if the singlets in $(55)$ sectors get
vev's, as mentioned above. Thus, generically only the standard
hypercharge will remain unbroken at low energies.

\bigskip

{\bf  ii) A three generation $SU(3)\times SU(2)_L\times
SU(2)_R\times U(1)$ model }

\bigskip

Let us consider now the $\IZ_3$ orientifold with twist action on CP
factors given by $N_0=2, N_1=N_4=2, N_2=7, N_3=3$, again as in the
previous example we choose to write explicitly the twist and
Wilson lines such that the $L-R$ model sector looks more
transparent:
\beq
V_9 \ = \
1/3(1,1,1,-1,-1,0,0,0,0,1,1,1,1,1,1,1) \eeq A quantized Wilson
line is added in the first complex plane given by:
\beq
W_9 \ = \
1/3(1,1,1,1,1,1,1,0,0,0,0,0,0,0,0,0) \eeq The gauge group from the
(99) sector will be $U(3)\times U(2)\times  U(2)
\times [SO(4)\times U(7)]$. Now the nine fixed points in the
first two complex planes split again into the same three sets
$(0,m)$, $(1,m)$ and $(-1,m)$ which have associated twists $V$,
$V+W$ and $V-W$ respectively. The corresponding value for $\Tr
\gamma_{\theta,9}$ are -4, -4 and -1. This means that we will only
need to add $5$-branes at the points $(-1,m)$ verifying    $\Tr
\gamma_{\theta,5}= -1$. The simplest option verifying this is
adding two 5-branes at each fixed point with $\gamma
_{\theta,5}=diag(\alpha , \alpha ^2)$. Again the $5$-branes are
trapped by tadpole conditions.

Let us now compute the spectrum which in this model is
particularly simple. One can check that the  $U(1)$ generator
defined by:
\beq
Q_{B-L}\ = \
(-2/3,-2/3,-2/3,-1,-1,-1,-1,0,0,...,0) \eeq is anomaly free and
can be identified with the standard $B-L$ symmetry of left-right
symmetric models. One finds chiral multiplets transforming under
the $SU(3)\times SU(2)_L\times SU(2)_R\times U(1)_{B-L}$ group:

{\bf (99):}

\beq
3(3,2,1,1/3)\ +\ 3({\bar 3},1,2,-1/3) \ +\ 3(1,2,2,0) \ +\ [
3(1,21)' +3(4,{\bar 7})'] \eeq These fields  include three standard
quark generations plus three sets of Higgs fields.

{\bf (59) :}

{}From the 5-branes at the $(-1,m)$ fixed points we get a $U(1)_5^3$ gauge
group. One then finds  chiral multiplets (in three copies, one per
fixed point) transforming like:
\beq
(1,2,1,-1)_q \ +\
(1,1,2,+1)_q \ +\ (3,1,1,-2/3)_{-q}+({\bar 3},1,1,2/3)_{-q} \ + \
[ (1,7)'_{q}\ +\ (4,1)'_{-q}] \eeq The subindex q indicates the charge under
the $U(1)_5^3$. These multiplets include three standard
generations of leptons and in addition three vectorlike sets of
color triplets.

{\bf (55) :}

For those $5$-branes sitting at fixed points $(-1,m)$ there is
just one singlet chiral field  $(1)_{2q}$ associated to the third
complex plane.

The massless spectrum of this model is remarkably simple. Notice
that, like in the previous model the extra colored objects may get
massive if the singlets in the $(55)$ sectors get a vev. Then the
massless spectrum contains just three standard generations and
three sets of Higgs fields.

\bigskip

{\bf iii) A three generation  $SU(4)\times SU(2)\times SU(2)$
model.}
\bigskip

Following similar lines we can consider the following twist action
on CP factors and quantized Wilson line in the first complex plane
for the $\IZ_3$ orbifold ($N_0=2, N_1=N_4=2, N_2=6, N_3=4$):
\beqa
 V_9\ = \ 1/3\, (1,1,1,1,1,1,0,0,0,0,1,1,1,1,1,1)\\ W_9\ = \
1/3\, (1,1,1,1,2,2,1,1,0,0,0,0,0,0,0,0)\nonumber \eeqa
The gauge group in the 99 sector is $U(4)\times U(2)\times
U(2)\times SO(4)\times U(6)$. The sectors feeling the action of
$V$ and $V+W$ do not need the addition of 5-branes (nor anti
5-branes) since the corresponding value for $\Tr
\gamma_{\theta,9}=-4$ cancels the tadpoles. On the other hand the
sector feeling $V-W$ has
 $\Tr \gamma_{\theta,9}=2$ and we need to add appropriate combinations
of 5-branes and anti 5-branes in order to satisfy tadpole
cancellation. A configuration of four 5-branes stuck at each of
the three fixed points $(-1,m)$ precisely satisfy the tadpole
constraint. In this case
$\gamma_{\theta,5}=diag(\alpha,\alpha,\alpha^2,\alpha^2)$ implying
a 5-brane gauge group $U(2)$ on each of the three fixed points. We
can identify the first factors of the 9-brane gauge groups as in
the Pati-Salam model: $SU(4)_c\times SU(2)_L\times SU(2)_R$. The
other factors may be considered as hidden sectors. The spectrum
comes again naturally in three families, in a self-explanatory
notation:

{\bf (99):}
\beq
3\, [(4,2,1)\ +\ (\bar 4,1,2)\ +\ (1,2,2) ]\ +\ 3\ [ (4,\bar
6)'\ + \ (1,15)'] \eeq
The observable part includes the complete three families of quarks
and leptons, as in the standard Pati-Salam model, plus three
copies of $SU(2)_L\times SU(2)_R$ Higgs fields.

{\bf (59):}

Since we have three identical copies of $U(2)$ on each of the
three fixed points, we will get three copies of:
\beq
[(4,1,1;2)_{-q}\ + \ (\bar 4, 1,1;2)_{-q}\ + \ (1,2,1;2)_q\ +
\ (1,1,2;2)_q]\ + \ [(4,1;2)'_{-q}\ + \ (1,6;2)'_q] \eeq
where the last entry on each particle represents its
transformation property under the corresponding $SU(2)$ and the
subindices the charge with respect to the $U(1)$. Notice that
again we have vector-like colored fields in this sector plus three
extra doublets of $SU(2)_L\times SU(2)_R$. However, with the usual
embedding of hypercharge into the Pati-Salam model their charges
are not standard.

{\bf (55):}

In this sector we only get a triplet plus two singlets under each
of the three $U(2)$ factors coming from the 5-branes. The triplet
may couple to the vector-like $4+\bar 4$ of the 59 sector to make
them massive. Also most of the $U(1)$ factors are anomalous, their
anomaly is cancelled by the Green-Schwarz mechanism and they do
not appear as low-energy gauge fields, only probably as remnant
global symmetries, as usual in these class of models. There is one
nonanomalous $U(1)$ surviving at low-energy. It is
\beq
Q\ = \ 1/2\ (1,1,1,1,-2,-2,2,2,0,\cdots ,0) \eeq
All other groups would decouple from the observable sector,
leading to a very simple string version of the Pati-Salam model. As
in the previous models, we will need an equal number (12) of anti
5-branes in the bulk to cancel the tadpoles, breaking
supersymmetry in a truly hidden sector.

\subsection{ Models with both  branes and antibranes stuck at fixed points}

One can also construct models in which both branes and antibranes
are both stuck at (different) fixed points. This possibility has
the attractive feature that provides an explanation to the absence
of (partial)  brane-antibrane annihilation without the need to resort to
other dynamical arguments. On the other hand these models are in
general non-supersymmetric already at the level of the massless
spectrum. There might be in general transitions between the
previous type models and the present ones as we will discuss
below. In fact we are going to build the models now by simply
replacing some of the $5$-branes by anti-$5$-branes in the
previous models in  such a way that the net total number of
$5$-branes charge is zero.

\bigskip
{\bf  i)  A non-SUSY three generation standard model}
\bigskip

Here on the same model of the previous section we can add a
different configuration of branes and anti-branes. The $(99)$
sector will be identical  and we will still add two $5$-branes at
each of the three fixed points $(-1,m)$ with $\gamma _{\theta ,
5}=(\alpha , \alpha ^2)$. Thus, at those fixed points we will have
three copies of fermions and complex scalars transforming like

${\bf (59) : }$

\beq
(1,2,-1/2)_q \ +\ (1,1,+1)_q \ +\ (3,1,-1/3)_{-q}+({\bar
3},1,1/3)_{-q} \ + \ [ (1,8)'_{q}\ +\ (4,1)'_{-q}] \eeq where the subindex
$q$ correspond to the $U(1)_5$ charges. We thus have here three
sets of right-handed leptons and three sets of left-handed leptons
(or $H_D$ Higgs superfields). {}From  the $(55)$ sector one gets one
singlet superfield $1_{2q}$ which has couplings to the vectorlike
quarks above. Thus the latter may become massive eventually if
this singlet gets a vev.

Now we will put two anti-$5$-branes at each of the three fixed
points $(0,m)$ with $\gamma _{\theta ,{\bar 5}}=(\alpha , \alpha
^2)$ in such a way that tadpole conditions are fulfilled. There
will be a $U(1)_{\bar 5}^3$ from the anti-$5$-branes. From $({\bar
5}9)$ sectors one gets matter fields in three copies transforming
like:

${\bf ({\bar 5}9) : }$

\beqa
Fermions &:& \ \ ({\bar 3},1,1/3)_r + (1,2,-1/2)_r +
(1,1,+1)_{-r}+(1,1,-1)_{-r}
                +[(1,{\bar 8})'_r+(4,1)'_{-r}]  \nonumber\\
Scalars  &:& \ \ ({\bar 3},1,1/3)_r + (1,2,-1/2)_r \ +\ [{\bar
8'}_r]\  + h.c. \eeqa
where the subindex $r$ refers to the
$U(1)_{\bar 5}^3$ charges. Notice that now this sector contains
the three right-handed $D$-quarks and three left-handed leptons,
as well as a set of vector-like right-handed leptons. From the
$({\bar 5}{\bar 5})$ sector one gets a couple of fermions
transforming like $1_{2r}$ and one singlet scalar $1_{2r}$, as well
as three extra singlet fermions  with negative chirality (the would be
gauginos of each $U(1)$ coming from the antibranes). The scalar $1_{2r}$
 has in general couplings with the vector-like pair of
right-handed leptons above which then may become massive
eventually.

Notice that the present model is explicitly non-supersymmetric
since some of the particles do not have SUSY-partners.
 On the
other hand the number of $5$-branes and anti-$5$-branes is the
same and are all stuck at the fixed points. Since there are only
$U(1)$'s coming from $5$ and anti-$5$-branes, there will be no
gauge group left from those since they are anomalous and will
become all massive in the usual way. In this sense this model is
simpler than its counterpart in section 4.1.
 The $U(1)$ symmetries will
remain however as effective global symmetries giving a flavour
structure to the model.

\bigskip

{\bf ii) A non-SUSY  $SU(3)\times SU(2)\times SU(2)\times U(1)$
model}
\bigskip

Again, on model ii) of previous section we distribute $5$-branes
and anti-$5$-branes in a different manner. The $(99)$ sector will
be identical but now on two of the fixed points (e.g., $(-1,0)$
and $(-1,1)$ ) we will add two $5$-branes with $\gamma _{\theta ,
5}=(\alpha , \alpha ^2)$ so that tadpoles cancel. These subsectors
will be  supersymmetric and we will get supermultiplets  in two
copies transforming like
\beq
(1,2,1,+1)_q \ +\ (1,1,2,-1)_q \ +\
(3,1,1,-2/3)_{-q}+({\bar 3},1,1,2/3)_{-q} \ + \ [ (1,7)'_{q}\ +\
(4,1)'_{-q}] \eeq where the subindices denote the $U(1)_5$ charge. Thus
from these two fixed points we get two lepton generations (plus
vector-like triplets which may become massive through couplings to
$(55)$ singlets). Now, on the remaining fixed point $(-1,-1)$ we
will add 4 anti-5-branes with $\gamma _{\theta , {\bar
5}}=(I_2,\alpha ,\alpha ^2)$ so that tadpoles cancel.
 There will be a
$(Sp(2)\times U(1)_{\bar 5})^3$ from the anti-$5$-branes. From
$({\bar 5}9)$
 sectors one gets
matter fields in one copy  transforming like:

${\bf ({\bar 5}9) : }$

\beqa
Fermions &:& \ \  (1,2,1,1;2)+(1,2,1,-1,;1)_r +
(1,1,2,-1;2) \nonumber \\
  & +& \ \ (1,1,2,1,;1)_r + ({\bar 3},1,1,1/3)_{-r}+ ({
3},1,1,-1/3)_{-r}\nonumber \\
         &+&      [ (1,\bar 7;2)'+(1,{7})'_r+(4,1)'_{-r} ]  \nonumber\\
Scalars  &:& \ \  [ ({\bar 3},1,1,1/3;2)+( 3,1,1,-1/3;2)+(4;2)' ] 
\nonumber
 \\
    &+&    [(1,2,1,-1,;1)_r+ (1,1,2,1,;1)_r +(7)'_r   + h.c.]
\eeqa
Here we get a third chiral generation of leptons and an
extra vector-like set of leptons. Again there are also vector-like
triplets which may become massive through scalars in the $({\bar
5}{\bar 5})$ sector. The latter has fermions  transforming like
$3(2)_{-r}\ +\ 2(1)_{2r}$ and  scalars  $3(2)_{-r}+(1)_{2r}$ which
indeed couple to the above vector-like triplets. It also has one
triplet of fermions of the $Sp(2)$ group as well as an extra singlet, both
 with negative chirality.

Notice that the spectrum of this model is supersymmetric except
for  one of the lepton generations.
 The number of $5$-branes and anti-$5$-branes is the
same and  all are stuck at the fixed points.
\medskip

{\bf  iii) A non-SUSY  $SU(4)\times SU(2)\times SU(2)$ model }

Here on the same model of the previous section we can add a
different configuration of branes and anti-branes. If we want all
extra branes stuck at orbifold fixed points, the simplest way
of distributing the branes and anti-branes is 4 branes on one
point, say $(-1,0)$,
 and two anti
5-branes at each of the other two points feeling the $V-W$ twist,
say $(-1,1), (-1,-1)$ . In those two points we have $\Tr
\gamma_{\theta,{\bar 5}}= diag(1,1)$ implying a group $USp(2)$ at
each of the two fixed points.

The spectrum of this model is as follows. The 99 sector is
identical to the one of the previous section, including the three
complete families of quarks and leptons. The 59 and 55 sectors
include only one copy of the one presented in the previous
section. Instead of the two other copies we have now the anti
5-brane sectors which are not supersymmetric.

 ${\bf (\bar 5 9):}$

On each fixed point we will have fermions (with opposite chirality
to the ones in the other sectors) transforming as:
\beq
[(1,2,1;2)\ + \ (1,1,2;2)]\ + \ [(1,\bar 6;2)'] \eeq where again
the last entry on each is the transformation under the
corresponding $Sp(2)$. The corresponding scalars will belong to the
following representations:
\beq
(4,1,1;2)+(\bar 4,1,1;2) + (4,1;2)'
\eeq
in two copies, one per fixed point.

 ${\bf(\bar  5\bar 5):}$

Here there are only two copies of a fermion  triplet one for each
 $Sp(2)$.

Notice that all non-abelian gauge anomalies cancel in a slightly
different manner to the model with only 5-branes stuck. Also, the
$(\bar 5 9)$ sector couples directly nonsupersymmetric matter to
particles charged under the observable gauge group, therefore
serving as direct mediators of supersymmetry breaking. This is the
main difference between the two models which otherwise are very
similar phenomenologically.

\subsection{A gravity mediated model with stuck branes and \\ antibranes}

The gravity mediated models discussed above have non-stuck
anti-branes somewhere in the bulk. These antibranes will in general be
attracted by the branes stuck at fixed points and the
configuration might be unstable. As we discuss in chapter 5, whether
this will be the case or not depends on the complete balance of
forces of the vacuum but anyway one would like to have some
example of gravity mediated model in which all branes and
antibranes are stuck. We will provide an example of this type
here.

The model we are going to discuss is a $\IZ_3$ orientifold with just
$3$-branes and anti-$3$-branes which is just a slight
(non-supersymmetric) modification of a supersymmetric model
presented in \cite{lpt}. The orientifold operation will be $\Omega
(-1)^FR_1R_2R_3$ which requires the presence of a net number of
$3$-branes minus anti-$3$-branes equal to 32 to cancel untwisted
tadpoles. Now, the model contains: a)  one fixed point (the
origin) which is also fixed  under the orientifold operation b)
other 26 fixed points under the $\IZ_3$ twist which are not fixed
under the $\IZ_2$ orientifold projection and c) 63 points which are
fixed under the orientifold operation but are not fixed under the
$\IZ_3$ action. At the origin, twisted tadpole conditions require
$\Tr \gamma_{\theta , 3}=-4$ whereas at the other 26 $\IZ_3$ fixed
points one has $\Tr \gamma _{\theta ,3} =0$.

We can now locate 11 $3$-branes at the origin with $\gamma
_{\theta ,3}=(1,\alpha I_5, \alpha ^2 I_5)$. The gauge group there
will be $SU(5)\times U(1)$ and there will be chiral multiplets
transforming like $3(10)_2+3({\bar 5})_{-1}$, i.e., three standard
$SU(5)$ generations. Notice that these 11 branes are stuck because
$3$-branes can only leave the origin in groups of 6 branes and
those remaining at the origin still have to obey the tadpole
conditions. Now we are going to add $21+3n$ $3$-branes and $3n$
anti-$3$-branes in such a way that all of them are trapped at some
points. In order to do that we recall that there is a simple way
to trap $3$-branes (or anti-$3$-branes) at some points. Indeed one
can locate one $3$-brane at one of the 63 fixed points under
$\Omega (-1)^FR_1R_2R_3$. In order for this configuration to be
invariant also under $\IZ_3$ we need to add one $3$-brane at each of
the two images of that point under $\IZ_3$. Now, these three
$3$-branes are stuck at those orientifold points because they can
only travel in groups of 6 in the bulk. So we can locate the
$21+3n$ $3$-branes in some of the 63 orientifold points available.
In the same way we can locate the $3n$ anti-$3$-branes  at some
other orientifold points. Now all branes and anti-branes in the
model will be stuck and all tadpole conditions are met. The
"observable" $SU(5)$ sector will only feel the presence of the
SUSY-breaking anti-$3$-branes from closed string exchange and the
effective field theory will be that of a gravity  mediated model.

Notice that one can stuck anti-$3$-branes at orientifold fixed
points also in the class of $\IZ_3$ models with $7$-branes and
$3({\bar 3})$ pairs mentioned above. Indeed, in this case the
orientifold operation is $\Omega (-1)^FR_3$ and we can locate
three anti-$3$-branes at the three points with coordinates
$(1/2,0)$, $(0,1/2)$ and $(1/2,1/2)$ in the 5-th and 6-th
dimensions. However we can only trap up to three anti-$3$-branes
in this way whereas we would need a minimum of 6 in the provided
examples. One could perhaps find this kind of models in a search
more systematic than the one we have made. The example provided in
this section is an existence proof of gravity mediated models with
all branes and anti-branes stuck.

\newpage
\begin{table}{}
\centering
\begin{minipage}{9.21in}
\scriptsize
\rotate[l]{
\vbox{
\begin{tabular}{|c|c|c|c|c|}
\hline
$(N_0,N_1,N_2,N_3,N_4)$ \hspace{0.3cm} &  Gauge Group \hspace{1.5cm}  & {\bf 99}
sector
 \hspace{ 1.5cm}& {\bf 59} sector  &  {\bf 55} sector \hspace{ 2cm} \\
$(n_a^i, m_a^i)$ & & & & \\
\hline\hline & & & &\\
 $(2,1,8,3,2)$ & $[SU(3)_c\times SU(2)_L\times U(1)_Y]$ &
$3[(3,2)+ (\bar 3,1)+ (1,2)$ & $3[(1,2)+ (1,1)+(3,1)+(\bar 3,1)
+(1,8)'+(4,1)' $ & $3(1)$ \\
$n_0^i=m_0^i=n_2^i=1$ & $\times U(1)^2\times [U(8)\times SO(4)]'$ & $
+(1,28)'+(4,\bar 8)']$ & $+(3,1)+(1,2)+2(1,1)+ (1,8)'+(4,1)'$
 & $3[3 
(2)+ 1]$    \\
 & $ \times [Sp(2)^3\times U(1)^6] $ &  & $+(\bar 3,1;2)+(1,2;2)+
(1,\bar 8;2)']$
 &  \\
\hline\hline
$(3,1,7,3,2)$ & $[SU(3)_c\times SU(2)_L\times U(1)_Y]$ & $3[(3,2)+
(\bar 3,1)+ (1,2)$ & $3[(\bar 3,1)+ (1,2)+ (1,2) +
(1,1)]+[(1,7)'+(6,1)'$ & $3(1)$ \\
$n_1^i=1, n_2^i=2$ & $\times U(1)^2\times [SO(6)\times U(7)]'$ &
$+(1,21)'+(6, \bar 7)']$ & $+(1,2;2)+(1,1;2)+(\bar 3,1,2) + (3,1;2) $ &
 $3[(1)+2(1)+(3)]$ \\
& $\times  [U(1)\times U(2)]^3$ & & $+
(1,7;2)'+(6,1;2)']$ &  \\
\hline\hline
$(1,1,9,3,2)$ &  $[SU(3)_c\times SU(2)_L\times U(1)_Y]$ & $3[(3,2)+
(\bar 3,1)+ (1,2)$ & $ 3[(\bar 3,1;2)+ (1,2;2) + (1,\bar 9;2)'+$ &
  \\
$m_0^i=n_1^i=m_1^i=1$ & $\times U(1)^2\times [SO(2)\times U(9)]'$ & $
+(2,\bar 9)'+(1,36)']$
& $(\bar 3,1) + (1,2)+(1,2) +(1,1)]+3[(2,1)'+(1,9)'$ &
$3[(2,1)+2(1,1)]$ \\
& $\times [Sp(2)^2\times U(1)]^3$ & & $+(3,1;2)+(1,2;2) +(1,1;2)]$ &
\\
\hline\hline
$(2,2,7,3,2)$ &  $[SU(3)_c\times SU(2)_L\times SU(2)_R]$ & $3[(3,2,1)+
(\bar 3,1,2)+ (1,2,2)$ & $3[(3,1,1)+(\bar 3,1,1)+(1,2,1)+(1,1,2)$ &
$ 3 (1)$ \\
$n_2^i=1$ & $\times U(1)^3\times [SO(4)\times U(7)]'\times U(1)^3$ & $+
(1,21)'+(4,\bar 7)']$ & $+(4,1)'+(1,7)']$
& \\
\hline\hline
$(1,2,8,3,2)$ &  $[SU(3)_c\times SU(2)_L\times SU(2)_R]$ & $3[(3,2,1)+
(\bar 3,1,2)+ (1,2,2)$ & $3[(3,1,1)+(1,2,1)+(1,1,2+2)+(2,1)'+(1,8)'$
& $3[3(2,1)+(1,1)]$ \\
$n_0^i=n_1^i=1$ & $\times U(1)^3\times[SO(2)\times U(8)]'$
& $+ (2,\bar 8)'+(1,28)']$ & $+ (\bar 3,1,1;2)+ (1,2,1;2)+ (\bar 8;2)'$ &  
\\
 $ m_0^i=m_1^i=1$ & $\times [Sp(2)\times U(1)]^6$ & & $+(\bar
3,1,1)+(1,2+2,1)+(1,1,2)+(2,1)'+(1,8)'$ &
$3[3(2,1)+(1,1)]$ \\
& & & $+(3,1,1;2)+(1,1,2;2)+(1,\bar 8;2)']$ & \\
\hline\hline
$(2,2,6,4,2)$ &  $[SU(4)_c\times SU(2)_L\times SU(2)_R]$ & $3[(4,2,1)+
(\bar 4,1,2)+ (1,2,2)$ & $3[(4,1,1;2)+(\bar
4,1,1;2)+(1,2,1;2)+(1,1,2;2)$ & 3[2 (1)+3] \\
$n_2^i=2$ & $\times [SO(4)\times U(6)]'\times U(2)^3 $ & $+(4,\bar 6)'+(1
,15)']$ &
$+ (4,1;2)'+(1,6;2)']$ & \\
\hline\hline
$(0,2,8,4,2)$ & $[SU(4)_c\times SU(2)_L\times SU(2)_R]$ & $3[(4,2,1)+
(\bar 4,1,2)+ (1,2,2)$ & $3[(\bar 4,1,1;2)+(1,2,1;2)+(\bar 8;2)'$ &
\\
$m_0^i=m_1^i=1$ & $\times [U(8)]'\times U(1)^3\times Sp(2)^6$ & $+ 28']
 $ &
$+(4,1,1;2)+ (1,1,2;2)+(\bar 8;2)']$ & \\
\hline \end{tabular}
\smallskip
\caption
{Models with
three generations and observable gauge groups like those of the Standard
Model,  left-right symmetric models and Pati-Salam models. The models
are determined by the values of $N_s, s=1,\cdots, 4$ and $n_a^i,
m_a^i$ $a,1=0,1,2$.
Only branes are trapped at the orbifold fixed points in these
cases. Representations coming from $SO(2N_0)\times U(N_2)$ are labeled
with a prime. Those coming from {\it each} of
the 5-brane groups are separated with a semicolon in the 95 sector. In
the 55 sector the two entries are the representations of the
$Sp(2m_a^i)\times U(n_a^i)$ group.}
\label{tabps1}}}
\end{minipage}
\end{table}

\newpage
\begin{table}{}
\centering
\begin{minipage}{9.205in}
\scriptsize
\rotate[l]{
\vbox{
\begin{tabular}{|c|c|c|c|c|}
\hline\hline  $(N_0,N_1,N_2,N_3,N_4) \hspace{ 1.cm}$
& {\bf 59} sector  & {\bf 55} sector \hspace{ 0.5 cm} & ${\bf \bar 5 9}$
sector &
${\bf \bar 5 \bar 5}$ sector  \hspace{ 2.7 cm}\\
  $n_a^i, m_a^i, q_a^j, p_a^j$ & & & & \\
\hline\hline & & & & \\
 $(2,1,8,3,2)$ &  $3[(1,2)+ (1,1)+(3,1)+(\bar 3,1) +(1,8)'+(4,1)'] $ &
 $3(1)$ &
$f_-:\, 3[(\bar 3,1)+(1,2)+ 2(1,1)+(1,\bar 8)' + (4,1)']$ & $f_+:\,  6(1),\
 f_-:\, 3(1)$ \\
$q_0^i=n_2^i= 1$ &  &  & $s:\, 3[(\bar 3,1)+(1,2)+(1,\bar 8)'+ h.c.]$
 & $s:\, 3(1)$   \\
\hline\hline
$(3,1,7,3,2)$ & $3[(\bar 3,1)+ (1,2)+ (1,2) +
(1,1)+(1,7)'+(6,1)']$ & $3(1)$ & $f_-:\, 3[(1,2;2)+(1,1;2)+ (1,7;2)']$
& $f_-:\, 3(3)$ \\
$n_1^i=p_2^i=1$ &  &  & $s:\, 3[(\bar 3,1;2)+(3,1;2)+(6;2)']$ & \\
\hline\hline
$(1,1,9,3,2)$ &   $ 3[(\bar 3,1;2)+ (1,2;2) + (1,\bar 9;2)']$ &
 & 
$f_-:\, 3[(\bar 3,1)+(1,2+2)+(1,1)+(1,9)']$ & $f_+:\, 3(1)\, f_-:\, 3(1)$  \\
$m_0^i=q_1^i=1$ &  & & $s:\ 3[(3,1)+(1,1)+(1,9)' + h.c.]$  & $s:\,
6(1)$ \\
\hline\hline
$(2,2,7,3,2)$ &  $2[(3,1,1)+(\bar 3,1,1)+(1,2,1)+(1,1,2)]$ &
$ 2 (1)$ & $ f_-:\,  (3,1,1)+(\bar 3,1,1)+(1,2,1)+(1,1,2)$ &
 $ f_+:\, 3(2)+2(1)$ \\
$n_2^{0,1}=p_2^2=q_2^2=1$ &  $+2[(4,1)'+(1,7)']$ & &
$+ (4,1)'+(1,7)'+(1,2,1;2)+(1,1,2;2)+(1,\bar 7;2)'$
& $f_-:\, (3)+(1)$ \\
& & & $s: (\bar 3,1,1;2)+(1,2,1)+(1,1,2)+7' +(4;2)'+ h.c.$ & $s: \ 3(2)+(1)$ \\
\hline\hline
$(1,2,8,3,2)$ &   $2[(3,1,1)+(1,2,1)+(1,1,2+2)+(2,1)'+(1,8)']$
& $2[3(2)+(1)]$ & $f_-:\, 3[(\bar 3,1,1)+(1,2+2,1)+(1,1,2)+8']$ & $f_+:\, 3(1)+
(1)$ \\
$n_0^{0,1}=m_0^{0,1}=1$ &   $+ 2[(\bar 3,1,1;2)+ (1,2,1;2)+ (8;2)']$ &
& $(3,1,1)+(1,2,1)+(1,1,2+2) + (2,1)'+(1, 8)'$ & $f_-:\, 3(1)+(1)$ \\
 $ q_1^j=q_0^2=1$ &  & & $s:\ 3[(3,1,1)+(1,1,2)+(1,8)']$ & $s:\
6(1)+2(1)$ \\
& & & $+ (3,1,1)+(1,1,2+2)+(2,1)'+(1,8)'$ & \\
\hline\hline
$(2,2,6,4,2)$  & $[(4,1,1;2)+(\bar
4,1,1;2)+(1,2,1;2)+(1,1,2;2)]$ & $2 (1)+3$ & $f_-:\ 2[(1,2,1;2)+(1,1,2;2)+
(1,\bar
6;2)'] $
 & \\
$n_2^0=2, p_2^{1,2}=1$ &
$+ 3[(4,1;2)'+(1,6;2)']$ & & $s:\ 2[(4,1,1;2)+(\bar 4,1,1;2)+(4,1;2)']$
& $f_-:\, 2(3)$ \\
\hline\hline
$(0,2,8,4,2)$ &  $2[(\bar 4,1,1;2)+(1,2,1;2)+(\bar 8;2)']$ & & $f_-:\ (\bar
4,1,1)+(1,2,1)+(1,1,2+2) + 8'$ & $f_+:\ 2[(1)+2(3)]$
\\
$m_0^{0,1}=m_1^{0,1}=1$ &
$+2[(4,1,1;2)+ (1,1,2;2)+(\bar 8;2)']$  & & $ (\bar
4,1,1;2)+(1,2+2,1;2)+(1,1,2;2) + (8;2)' $ & $f_-:\, (3)+(3)$ \\
$q_0^2=q_1^2=2$ &  &  & $s:\ 2(4,1,1)+(1,2,1)+(1,1,2)+2(8)'+h.c.$ & 
$s:\ 2[2(1)+(3)]$ \\
\hline
\end{tabular}
\smallskip
\caption{Models with three
generations and observable gauge groups like those of the Standard
Model,  left-right symmetric models and Pati-Salam models. The models
are determined by the values of $N_s, s=1,\cdots, 4$ the number of
trapped 5-branes  $n_a^i,
m_a^i$ $a,1=0,1,2$, and the number of trapped anti 5-branes
$p_a^j,q_a^j$. Gauge groups and $99$ sectors are as in table 1, except
for the groups coming from the anti 5-branes which are
$\Pi_{a,j} Sp(2p_a^j)\times U(q_a^j)$.
 Representations coming from $SO(2N_0)\times U(N_2)$ are labeled
with a prime. Those coming from {\it each} of
the 5-brane or anti 5-brane groups are separated with a semicolon
in the $59$ and $\bar 5 9$ sectors. In the non-supersymmetric sectors
we specify the fermion spectrum (including chirality) and the scalars also.}
\label{tabps2}
}}\end{minipage}
\end{table}
\clearpage
\newpage

\section{Scales and dynamics of the vacua}
\subsection{Lowering the string scale}

The Type I vacua in previous sections are non-supersymmetric. The
scale of SUSY-breaking is thus essentially the string scale $M_{string}$.
In the second class of models in which some quark-lepton sectors
are non-SUSY already at the level of the spectrum, it is clear
then that the string scale cannot be much higher than say $M_s
\leq  1-100$ TeV, otherwise all scalars will get masses in loops
and we will have to face the usual gauge hierarchy problem. In the
first class of models with anti-$5$-branes in the bulk and hidden
from $9$-branes, SUSY-breaking is hidden from the observable world
coming from $(99)$ and $(59)$ sectors. Thus SUSY-breaking will be
only felt in a suppressed manner and one would guess that there may
appear SUSY-partner masses $M_{SB}$ of order
$M_{SB}=M_{string}^2/(M_{Planck})$ as usual in hidden sector models.
Thus,
if we want $M_{SB}$ not to exceed the weak scale   $M_W$ so that
the solution to the hierarchy problem is not spoiled we must
require:
\beq
M_{string} \ \leq \ \sqrt{M_W M_{Planck}} \ \propto
10^{11} \ GeV \eeq Thus in any of the schemes here considered we
need to lower the string scale well below the Planck mass, at
least down to the intermediate scale.

It is by now well known that one can lower the string scale in
generic Type I vacua \cite{witten,lykken,antoniadis}.
 In the $\IZ_3$ case at hand it is more
appropriate to perform a T-duality transformation along the third
compact complex plane and work with $7$-branes instead of 9-branes
and $3$-, anti-$3$-branes instead of $5$-, anti-$5$-branes. The
orientifold operation will be now $\Omega (-1)^{F}R_3$, where
$R_3$ is the reflection operation with respect to the third
complex plane.
 The world-volume of the 7-branes
includes the first two complex dimensions and they are located at
the origin $X_3=0$ in the third complex plane. As in the case of
$9$-branes, one can add Wilson-lines in the first two complex
planes and one can distribute $3$-branes or anti-$3$-branes
located anywhere in compact space. The distribution of $3({\bar
3})$-branes
 and additions of Wilson lines in the first two complex planes
is subject to the tadpole constraints:
\beq
\Tr ({\cal W})^k
\gamma_{\theta,7} + 3(\Tr\gamma_{\theta,3,L}-
\Tr\gamma_{\theta,\bar{3},L}) = -4 \label{tadpz3} \eeq where $L$
labels the 9 fixed points in the first two complex planes. Thus
the model so obtained has exactly the same massless spectrum and
interactions as the original one. Now, if we locate the
anti-$3$-branes at points with $X_3\not=0$, they will have no
overlap with the 7-branes and they will act as a SUSY-breaking
hidden sector for the $(77)$ and $(73)$ massless fields. This is
the T-dual of the  class of models considered in section 4.1.

Let us consider the gauge couplings and scales in this class of
orientifolds. The Planck mass $M_p$ is related to the compact
scales $M_i$, $i=1,2,3$ and 10-dimensional Type I dilaton $\lambda
$ by (see e.g. ref.\cite{imr}) :
\beq
M_p\ =\ { {2\sqrt{2} M_s^4}\over {\lambda M_1M_2M_3} }
\eeq and the gauge couplings of the gauge groups from $7$-branes
and $3$-branes are
\beq
\alpha _7\ =\ { {\lambda M_1^2M_2^2 }\over
{2M_s^4} } \ \ ;\ \ \alpha _3 \ =\ { {\lambda }\over 2} \eeq Thus,
irrespective of the string scale one has:
\beq
 M_p\ =\
{{\sqrt{2}}\over {\alpha _7}} { {M_1M_2}\over {M_3} } .  \eeq By
making $M_3$ much smaller than $M_{1,2}\propto M_s$, one can
obtain agreement with the measured values of $M_p$ and $\alpha
_7$. In particular, one can set $M_1=M_2=M_s$, which leads to
$\alpha _7 =\alpha _3 =\lambda /2$. One can now take the string
scale at the intermediate scale $M_{1,2}=M_s=10^{10}$ GeV
 by choosing $M_3\propto 1$ TeV. In the same way, one can
take $M_s =M_{1,2}= 1$ TeV  by taking $M_3=10^{-3}$ eV.

One concludes that the string scale, which in this case coincides
with the SUSY-breaking scale, may be arbitrarily lowered by making
sufficiently large the third compact dimension. Notice that in the
present scheme the first two compact dimensions are small, of
order the string scale, whereas the third dimension is large.
There may be a possible dynamical explanation for this asymmetry
in the second class of models discussed in the previous sections
where branes and anti-branes are stuck at fixed points in the
first two complex compact dimensions. Indeed, as emphasized in
ref.\cite{au}, in this case one expects that the brane-anti-brane attraction
will generate a potential which will tend to shrink the compact
radii $r_{1,2}$. This force would not be present for the third
compact dimension which can then dynamically grow.

\subsection{Stability and Effective Potentials}

Another issue we need to address is the stability of these
 configurations.
 We know
that contrary to most previous considerations of brane/anti-brane
systems, there are no tachyons  that could destabilize these
constructions \cite{au} .
 Furthermore, some or all of
the branes must be trapped in the orbifold fixed points in order
to satisfy tadpole cancellation. For instance, for
 the situation with only branes stuck
at the fixed points, there will be the same number of anti-branes
living on  the bulk and one may wonder why they do not approach
the trapped branes and annihilate. In fact the twisted tadpole
cancellation conditions in general forbid this complete
annihilation process because some residual branes or antibranes
have to remain in the fixed points to cancel the tadpoles. Thus
the annihilation can only  be partial.

In this connection it is interesting to study
 the possible relation between the models
of section 4.1 and those of section 4.2. One can imagine starting
in a configuration with branes stuck at the fixed points and
anti-branes moving freely on the bulk, like  those of section 4.1.
Eventually the right number of anti-branes would get attracted to
the right fixed points, annihilating some of the branes at those
points in a manner consistent with tadpole cancellation.
 For instance in the $SU(3)\times SU(2)\times SU(2)$ model, we
may start with the configuration with three pairs of stuck
5-branes and six anti 5-branes on the bulk. All of these anti
5-branes may go to one single fixed point, annihilate the two
5-branes there, leaving four anti 5-branes and becoming the
corresponding model of section 4.2. In the $SU(4)\times
SU(2)\times SU(2)$ model however, the twelve anti 5-branes would
have to go, six to one point and six to the second point in order
to keep tadpole cancellation. It is interesting to ask which brane
configuration is dynamically preferred, if any, since both seem to
be connected in some way. This seems to be a complicated dynamical
question since several dynamical effects seem to be competing.
 Usually since branes and
anti-branes carry opposite Ramond-Ramond charge they tend to
attract each other \cite{au} .
 However,  the presence of the stuck branes also
induces a repulsive interaction between them and the anti-branes,
due to the total positive brane tension,  similar to the effect of
a positive  cosmological constant in inflationary models, for
instance. This extra contribution to the vacuum energy corresponds
actually to an effective potential for the dilaton and the moduli
fields. In general the exact form of this potential is something
beyond our control, since supersymmetry has been broken, there are
all sorts of radiative corrections that can contribute as well as
standard non-perturbative effects
\footnote{For some recent discussions of radii stabilisation
in the context of brane models see refs.\cite{radii} .}

In the case in which both branes and anti-branes are stuck at
fixed points (second class of models in the previous sections)
naively one could argue that there are competing effects, some of
them trying to shrink the first two compact dimensions and other
trying to expand them, as we now explain.
 For the models
constructed in terms of 9 and 5-branes, the tension of the
5-branes and anti 5-branes is proportional to $1/\lambda$, with
$\lambda$ the string coupling. Upon compactification the tension
will generate a term in the action proportional to
$1/M_3^2\lambda$ where $M_3$ is the compactification scale of the
third complex direction which the 5-brane  is wrapped around. This
is precisely the real part of the so-called $T_3$ field of type I
compactifications. Therefore we can see that the brane tension
induces a linear potential for one of the moduli fields. In the
dual version in terms of 7 and 3-branes, which is more appropriate
to discuss a lowering of the string scale,
 this term is only proportional to
$1/\lambda$ which is the corresponding dual $T_3$.
 Recall that in the latter  brane configuration,
 the moduli fields are given by
the following expressions \cite{imr}:
\beqa ReS\, &  = & \, \frac{2 M_s^4}{\lambda M_1^2 M_2^2}\ \ \ ; \
\ ReT_3\ = \frac{2}{\lambda} \
\\ \nonumber
ReT_1\  & =  & \ \frac{2M_s^4}{\lambda M_2^2 M_3^2} \ \ \ ; \ \
ReT_2\ = \ \frac{2M_s^4}{\lambda M_1^2 M_3^2}
 \eeqa
The gauge couplings  are given by ${\rm Re} S$ for the groups on
the 7-branes and ${\rm Re} T_3$ for the groups on the 3-branes.
Therefore perturbative corrections in the 7-branes field theory
 would go like $S^{-n}$ whereas non-perturbative corrections would
depend on $e^{-S}$.
 Therefore $S$ seems to have
the standard runaway behaviour towards large values
 whereas $T_3$ has the linear
contribution from the brane tension that tends to stabilize it at
small
 values. Thus these two effects tend to drive
$M_2^2M_3^2=\frac{{\rm Re}T_3}{{\rm Re}S}$ towards {\it  small }
values. On the other hand there is in addition
 the contribution to the vacuum energy from the
brane/anti-brane attraction, that, as argued above, tends to drive
$M_2^2M_3^2$ to {\it large } values. Thus there seems to be two
competing effects in opposite directions. This could lead to two
possible outcomes: i) The competing effects lead to stable
minimum with fixed $M_{1,2}$ and $\lambda $ or ii) At a given
moment it could be more energetically favorable that anti-branes
pop out from the fixed points to the bulk  and only branes
(satisfying tadpole cancellations) remain there. This would mean a
transition from the type of  models of section 4.2 to those of
section 4.1. Without more dynamical information it is difficult to
say which
 type of configurations,
either with trapped anti-branes at fixed points or with antibranes
away in the bulk is preferred from  the effective potential point
of view.

All these considerations need further scrutiny in order to be able
to extract concrete implications about the effective dynamics of
these new brane configurations. We have the advantage over
previous discussions that the models are explicit and the
effective action is in principle computable, at least in some
approximation scheme. There is no need to say that after fixing
the dilaton and moduli fields, we would have to be exposed to the
standard cosmological constant problem, as in any other treatment
of supersymmetry breaking.

\subsection{ Yukawa couplings}

This is not the place to make a detailed phenomenological analysis
 of Yukawa couplings which we leave to future work.
It is however interesting to present the general structure of
Yukawa couplings in this class of theories, which turn out to be
quite different to similar heterotic orbifold models.
 The following renormalizable couplings are generically
present:

{ \it i) $(99)(99)(99)$ couplings }

There are superpotential couplings of the form :
\beq
\phi^{99}
_i\phi^{99} _j \phi^{99} _k \ ,\ \  i\not= j\not= k\not= i \eeq
where $\phi^{99} _i$, $i=1,2,3$ are charged chiral fields in the
$(99)$ sector associated to the complex plane $i$. These type of
couplings give rise for example to quark Yukawa couplings in the
$SU(3)\times SU(2)_L\times SU(2)_R\times U(1)$ models of chapter
4. These couplings are totally analogous to the ones present for
untwisted particles in perturbative heterotic orbifolds. The
couplings $(55)^3$ have a completely analogous structure.

{\it ii) $(59)(99)(95)$ couplings}

The worldvolume of $5$-branes includes the third complex plane.
Thus there are superpotential couplings of the form:
\beq
\psi
^{59}_a\psi ^{59}_a \phi^{99}_3 \eeq where $a$ label the fixed
point where the $5$-brane is localized. Notice that only the
$(99)$ chiral fields from the third complex plane have these
couplings. In addition these couplings are diagonal in the $a$
label, i.e., there are no renormalizable couplings involving
different fixed points. These type of couplings give rise for
example to standard lepton  Yukawa couplings in the $SU(3)\times
SU(2)_L\times SU(2)_R\times U(1)$ models of chapter 4.

{\it iii) $(55)(59)(95)$ couplings}

In a similar manner there are superpotential couplings of the form
\beq
 \psi ^{59}_a\psi ^{59}_a \phi^{55}_{3,a} \eeq in which again
$a$ labels the fixed point. Only the $(55)$ chiral fields in the
third complex plane appear in the coupling. For example, There is
a coupling of this type between a singlet $(1)_{2q}$ in the $(55)$
sector and the coloured pair  $(3,1,1,-2/3)_{-q}+({\bar
3},1,1,2/3)_{-q}$ in the same left-right symmetric model  of
chapter 4. In general the presence of this type of
couplings tends to reduce the number of extra vector-like
fields in the massless sector once scalars in the $(55)$
sectors get vevs.

{\it iv) $(99)({\bar 5}9)(9{\bar 5})$ couplings }

The  $({\bar 5}9)$ sectors are not supersymmetric so that some
SUSY partners are missing. Again, only the $(99)$ fields in the
third complex plane have this type of couplings. There are Yukawa
couplings of type:
\beqa
  &  \psi ^{{\bar 5}9}_a\psi ^{{\bar 5}9}_a \phi^{99}_{3,a} &
\\ \nonumber
&  \phi ^{{\bar 5}9}_a\psi ^{{\bar 5}9}_a \psi^{99}_{3,a} & \eeqa
in which now $\phi $ denotes a scalar and $\psi $ a fermion. Examples
of these are the lepton Yukawa couplings in the left-right
symmetric model in section 4.2.

{\it v) $({\bar 5}{\bar 5})({\bar 5}9)(9{\bar 5})$ couplings }

In a similar manner there are Yukawa  couplings of the form
\beqa
 & \psi ^{{\bar 5}9}_a\psi ^{{\bar 5}9}_a \phi^{{\bar 5}}_{3,a}
& \\ \nonumber & \phi ^{{\bar 5}9}_a\psi ^{{\bar 5}9}_a
\psi^{{\bar 5}}_{3,a} & \eeqa in which again $a$ labels the fixed
point and $\phi $($\psi $) denotes a scalar(fermion). Only the
$({\bar 5}{\bar 5})$  fields in the third complex plane appear in
the coupling. For example, there is a coupling of this type
between a singlet scalar $(1)_{2r}$ in the $({\bar 5}{\bar 5})$
sector and the coloured pair of fermions
 $(3,1,1,-2/3)_{-r}+({\bar 3},1,1,2/3)_{-r}$ in
the  left-right symmetric model  of section 4.2.

As we mentioned above, the structure of the $(55)^3$ couplings is
analogous to that of the $(99)^3$ couplings. Something similar
happens for the couplings of type $({\bar 5 }{\bar 5})^3$,
although, of course in this sector the spectrum is not
supersymmetric. Still the only Yukawa couplings which do not
vanish are those involving fields in three different complex
planes. Concerning the sizes of the couplings, $(99)^3$,
$(59)(99)(95)$ and $(99)({\bar 5}9)(9{\bar 5})$ couplings are
proportional to $(ReS)^{-1/2}$, $S$ being the complex dilaton
defined above. $(55)^3$, $({\bar 5 }{\bar 5})^3$, $({\bar 5}{\bar
5})({\bar 5}9)(9{\bar 5})$ and $(55)(59)(95)$ couplings are
proportional to $(T_3)^{-1/2}$ as defined above.

It is interesting to remark that all models present a flavour
structure coming from the different $5$-brane groups. In the case
in which $5$ branes give only rise to $U(1)$'s, we have already
seen how those $U(1)$'s are broken by the Green-Schwarz mechanism
and effective global $U(1)$'s effectively persist. These $U(1)$
symmetries (and possibly larger $5$-brane gauge symmetries) may
play an important role in ensuring sufficient proton stability. We
postpone a through analysis of this question for future work.

\subsection{Gauge coupling unification}

An analysis of this question should be done on a model by model
basis and such an analysis goes beyond the scope of this paper.
 We would
like to point out however that in some of the
models discussed above there is  a tendency to get
gauge coupling unification at the
intermediate scale.
 Specifically let us consider the SM orientifold of
 section 4.1. Notice first that the embedding
of the $U(1)$ hypercharge  does not have the canonical
normalization with $g_1^2/g_2^2 =3/5$ at the string scale. One
rather has $g_1^2/g_2^2 =3/11$. Then the one-loop renormalization
group running from the string scale $M_s$ to the weak scale $M_Z$
gives for the weak angle and the QCD couplings at the weak scale:
\beqa
\sin^2\theta _W(M_Z)\ =&\ {3\over {14}}(1\ +\ {{11\alpha
(M_Z)}\over {6\pi }}\ (b_2-{3\over {11}}b_1)\ \log({{M_s}\over
{M_Z}})\ ) \nonumber  \\
{1\over {\alpha _3(M_Z)}}\ =&\ {3\over
{14}}({1\over {\alpha (M_Z)}}\ -\ {1\over {2\pi }}\
(b_1+b_2-{{14}\over 3}b_3)\ \log({{M_s}\over {M_Z}})\ )
\label{senos}
\eeqa
In order to perform the running we only have
to include the particles which remain massless below $M_s$. In the
particular models at hand that depends on whether certain singlet
fields in the $(55)$ sectors get vevs and give masses to some
vector-like particles. We will consider the most generic case in
which indeed those singlets get vevs of order the string scale.
 Let us assume then that,  the three sets of chiral
 multiplets $ (3,1,-1/3)_{-q}+({\bar
3},1,1/3)_{-q} $ and also the vector-like right-handed
leptons from the $(0,m)$ fixed points
get a mass of order $M_s$ due to the couplings of
these  fields to the singlet scalars in the $(55)$ sectors.
 Then below
the string scale the $\beta $-functions of the $SU(3)\times
SU(2)\times U(1)$ interactions are respectively $b_3=0$ , $b_2=6$
and $b_1=18$. Plugging this in the above renormalization group
formula one obtains:
\beq
M_s\ =\ 1.4 \times 10^{12} \ GeV \ \ ;\ \
sin^2\theta _W(M_Z)\ =\ 0.239
\eeq
where we have used $\alpha_3(M_Z)=0.12$.
Taking into account that in this simple estimate
we have just made  a one-loop
computation and  we have  also  assumed all light
thresholds at $M_Z$ and all heavy thresholds
close to $M_s$, these are not unreasonable values,
given the uncertainties. In particular, not all
fields which get masses through the vevs of
$(55)$ scalars need to have identical masses
and decouple at the same scale. For example,
one can check that if the 3 sets of right-handed
vector-like leptons have masses below $M_s$,
both the values obtained for $sin^2\theta _W$ and $M_s$
decrease accordingly.

In models like those of section 4.2 which are explicitly non
supersymmetric, the string scale must be at most of order 1-100
TeV to avoid the hierarchy problem. Thus the usual field theory
logarithmic running will not be in general sufficient to achieve
unification at such low string scale. Other approaches like
perhaps  those
of refs. \cite{mirage,halyo,bachas2,adm}
(see also \cite{bajogut,gr,kaku3,ddg} )
 could  be relevant. Meanwhile one must admit
that models with only branes at fixed points like the ones in
section 4.1 seem to be able to easily accommodate gauge coupling
unification as long as the string scale is of order of the
intermediate scale which, as we mentioned above, is the natural one if
we have gravity mediated SUSY-breaking.

\section{Final comments and outlook}

In this paper we have constructed the first semirealistic string
vacua  using D-brane techniques. We have showed that the addition
of brane-antibrane pairs on otherwise $N=1$, $D=4$ supersymmetric
Type I models is a quite versatile technique in order to produce
tachyon-free and RR tadpole-free string vacua resembling the
standard model of particle physics (or some left-right symmetric
generalizations of it).

 The vacua constructed are
non-supersymmetric but the SUSY-breaking effects may be felt in a
suppressed manner in some subsectors of the theory. For example,
the sector identified with the SM
 may
be separated in transverse compact space from the SUSY-breaking
anti-branes in such a way that only closed strings can transmit
SUSY-breaking to the observable physics world. This would
correspond to a standard gravity mediated scenario in which
SUSY-breaking has to happen at the intermediate scale
$\sqrt{M_WM_{Planck}} \propto 10^{11}$ GeV. These are the class of
models presented in sections 4.1 and 4.3. Since in this class of
theories the SUSY-breaking scale is the string scale one would
have to identify $M_s$ with the intermediate scale also
\footnote{In this setting only closed string states
living in the bulk like the dilaton $S$ and complex moduli $T_i$ 
couple simultaneously to branes and antibranes. It is 
thus reasonable  to expect that
the SUSY-breaking effects transmitted to the $(99)$ and $(59)$ 
observable sectors may be parametrized in terms of vev's for the
auxiliary fields of $S$, and $T_i$, as propossed in refs.
\cite{spurion,imr}.}.  We have
explicitly shown how in this class of theories one can indeed
lower the string scale to these values and still maintain
consistency with the known values of Planck mass and perturbative
gauge couplings. We have also shown  that in some of the specific
examples presented the standard model couplings do indeed join at
the intermediate scale. This comes about because of the presence
in these models of extra vector-like lepton generations or
Higgs doublets.

In other constructions   SUSY-breaking may be transmitted to the
observable world either by gauge mediation (as in the example in
the appendix) or even at the tree level, i.e., one can get some
open string sectors with non-SUSY spectra. The latter is the case
of the models presented in section 4.2. In this case the
SUSY-breaking  (and thus the string scale) should not be above say
1-100 TeV if we want to avoid the standard hierarchy problem.

We have presented specific examples of string vacua with three
generations and the gauge group of the standard model or some of
their left-right symmetric extensions but one can construct many
more along similar lines.
 The models constructed provide for specific examples
in which recent ideas involving a string scale well below the
Planck scale may be tested.
 Their massless spectrum is quite simple compared with
analogous perturbative heterotic models constructed in the past.
Thus, for example, the left-right symmetric model of section 4.2
contains just three quark-lepton generations and three sets of 
Higgsses in its massless spectrum. 
We have concentrated on the $\IZ_3$ orientifold because three
generations are easily obtained in that case but it should be
possible to construct many more examples using other orientifolds.

We have not presented a detailed phenomenological analysis of the
specific models presented which we leave for future work. We have
however discussed the general structure of $U(1)$ anomaly
cancellation and Yukawa couplings which are the required
ingredients for such a phenomenological analysis. In general there
are several anomalous $U(1)$'s whose anomalies are cancelled by
the generalized Green-Schwarz mechanism discussed in ref. \cite{iru}.
There are associated Fayet-Iliopoulos terms which are controlled by
the blowing-up moduli of the orbifold. Generically only the
standard hypercharge $U(1)$ remains unbroken down to low energies
in the explicit SM orientifolds constructed. The broken $U(1)$
symmetries lead to some flavour structure for the models which
deserve further analysis since they very much determine the
structure of Yukawa couplings and hence questions like fermion
textures and proton stability.

It is perhaps instructive to compare this class of realistic  Type
I vacua with those first constructed in ref.\cite{iknq}
 in the context of
$\IZ_3$ heterotic perturbative orbifolds. Models like the ones we
have constructed here do not have a perturbative heterotic dual
since, e.g., the gauge shifts $V$ and Wilson lines $W$ used do not obey
the standard modular invariance constraints. However the models in
refs.\cite{iknq}  have a number of similarities (and also important
differences) with the ones built here. One of the similarities is
that the models constructed with both techniques have a tendency
to give rise to the presence of extra vector-like sets of leptons
at low energies. However, important aspects like the structure of
Yukawa couplings are totally different. In general the Type I
constructions are much simpler in most  respects: 1)
Only the addition of one Wilson line  is required in Type I
whereas two Wilson lines (leading to a proliferation of twisted
sectors and massless states) are required in the heterotic models.
2) In the heterotic models the necessary presence of a
non-vanishing (dilaton-dependent) Fayet-Iliopoulos term makes also
the identification of the correct perturbative vacuum quite
cumbersome, making a detailed analysis of the effective
field-theory D-flat directions necessary. In the Type I case models
one can put in principle the FI-terms to zero and no complicated
analysis of flat directions is needed.
 3) In the Type I case one can lower the string scale down to the
intermediate scale and achieve gauge coupling unification.
 The possibility of lowering the scale is not available
in perturbative heterotic models and this makes difficult to
achieve gauge coupling unification. 4) In the Type I models here
constructed SUSY-breaking (in a hidden sector or not) is a
built-in property of the models. In perturbative heterotic models
one relies on the possible existence of some hierarchy-generating
mechanism like gaugino condensation in order to give rise to
SUSY-breaking.

An interesting difference between these models and the perturbative
heterotic models is the rank of the gauge group. In perturbative
models, the rank cannot exceed 22 and in orbifold models
at generic values of the moduli space of the six dimensional compact space,
the gauge group has rank 16. In the present models, the rank of
the
group coming from the 9-branes is also  16. On the other hand,
 the rank of the
groups coming from the 5 and $\bar 5$ branes may be very large,
depending on the values of $N_s$ and it seems to be  unlimited.
 We may
easily see this by considering the tadpole conditions (\ref
{tad} ). If for instance the left hand side of those
equations vanishes, as in the supersymmetric case,
 we can have, say $n_1^i=2m_1^i=q_2^i=2p_2^i=2K$
with $K$ arbitrarily large and all equations will be satisfied. This
seems to imply  that as long as the number of branes and anti branes is
equal there is a lot of freedom to satisfy the tadpole conditions and an
arbitrary number of branes and anti branes can be added satisfying
those conditions, the corresponding rank of the gauge groups increases
linearly with $K$ and we have no bound on this value.
 It was known
previously that non perturbative string vacua can have rank much
larger than 22. Here, we seem to have explicit models with an
unbounded 
rank,
bringing the difference with perturbative vacua to the extreme.
However we expect these higher rank models to be unstable and `decay'
to the lowest rank models, the reason being the following: if in the
example above we can satisfy the tadpole equations with say
$n_1^i=m_1^i=0$. The addition of an arbitrary number
 of branes that still keeps the conditions satisfied
($n_1^i=2m_1^i=2K$) is not
necessary for tadpole cancellation therefore
 the corresponding extra branes (and
anti-branes) do not need to be trapped at the fixed points. We then
expect that those configurations will decay to the one with the minimum
number of branes required by  the tadpole conditions, which are the
only ones forced to be trapped at the fixed points. All the extra
branes and anti-branes may annihilate each other in the bulk.
 Recall that
the explicit models we have presented require the minimum number of
branes and anti branes.

One open question that one should study more carefully is that
 of the stability of this class of vacua. As we
remarked above, these models are free of tachyons both in the bulk
and on the branes. Some of the branes and/or anti-branes are stuck
at some of the fixed points of the underlying orbifold due to the
tadpole cancellation conditions  and hence they are not free to
travel to the bulk and give rise to complete brane-antibrane
annihilation. The existence of these trapped branes may lead to
dynamical effects tending to shrink some of the compact
dimensions and no others as discussed in the text.

These are first steps in the search for realistic string vacua
with the D-brane techniques. Much work remains to be done in this
direction both from the theoretical side, finding new classes of
vacua and addressing the important dynamical issues involved in
this class of theories, as well as from the phenomenological side,
studying the viability of specific examples as well as extracting
general properties which could be generic. We hope we have
convinced the readers of the interest of such a search.

\bigskip

\bigskip

\centerline{\bf Acknowledgements}
We give special thanks to Angel Uranga for many comments
and suggestions while this work was being carried out.
This work has been partially
supported by CICYT (Spain), the European Commission (grant
ERBFMRX-CT96-0045), the John Simon Guggenheim foundation and PPARC.
G.A work is partially supported by APCyT grant 03-03403. 
 L.E.I. also thanks CERN's Theory Division where part of this 
work was done.

\newpage

\section{Appendix}

\subsection{A model with gauge-mediated supersymmetry breaking}

The class of models with $9$-branes and $5$-branes discussed in
chapter 3 allow also for the construction of simple models in
which the supersymmetry breaking effects originated from
anti-$5$-branes is transmitted to an ``observable'' world of
$5$-branes through loops involving gauge interactions from the
$9$-branes living in the bulk. In particular, the SUSY chiral
multiplets in $(59)$ sectors couple through $(99)$ sector gauge
interactions to the non-SUSY $({\bar 5}9)$ sector.

Consider as an example the following $\IZ_3$ orientifold. We take
the 32 $9$-branes with twist on CP factors given by:
\beq
\gamma
_{\theta , 9}\ =\ diag(I_{18}, \alpha I_{7} , \alpha ^2 I_7) \eeq
Now, since $\Tr \gamma _{\theta , 9 }=11$, tadpole cancellation
conditions at each of the 9 fixed points in the first two complex
planes require:
\beq
\Tr \gamma _{\theta , 5}\ - \ \Tr \gamma
_{\theta , {\bar 5}} \ =\ -5 \eeq Now, we will locate   10
$5$-branes at each of 4 fixed points with:
\beq
\gamma _{\theta ,
5}\ =\ diag( \alpha I_{5} , \alpha ^2 I_5) \eeq and at each of the
remaining 5 fixed points 8 anti-$5$-branes with
\beq
\gamma
_{\theta , {\bar 5}}\ =\ diag(I_6, \alpha  , \alpha ^2) \ . \eeq
The complete  gauge group is:
\beq
[U(5)^4]_5\times [ SO(18)\times
U(7)]_9\times [(Sp(6)\times U(1))^5]_{\bar 5} \eeq We can now
identify one of the four $SU(5)$'s with the observable physical
world. In the $(55)$ sector one has charged chiral multiplets
transforming like: $2(10)+15$ under $SU(5)$. Now, from the $(59)$
sector we will get supermultiplets transforming like
 $(18,1;{\bar 5})$ $+(1,7;5)$ under $SO(18)\times SU(7)\times SU(5)$.
Altogether the $SU(5)$ charged spectrum will contain $2(10+{\bar
5})+ (15+ 9({\bar 5}))$ $+7(5+{\bar 5})$. This is an anomaly-free
spectrum with two standard plus one exotic $SU(5)$ generations.
This spectrum is totally supersymmetric but it will get non-SUSY
corrections in loops coming from their couplings of $SO(18)\times
U(7)$ gauge bosons    to the non-SUSY $({\bar 5}9)$ sector.
Although this particular model is not very realistic, it
exemplifies how in this type of vacua one can indeed obtain models
with gauge mediated SUSY breaking as advertised in the text.

\subsection{A variant SM }

One can construct many variations on the models presented in
sections 4.1 and 4.2. The $SU(3)\times SU(2)\times U(1)$ model of
4.2 has the right-handed quarks in a non-SUSY sector of the
theory. This can be easily modified and e.g., one can get a
variation of the model in which only the leptons have a non-SUSY
spectrum. Let us first start with the model with only $5$-branes
at the fixed points and then we will discus the model with both
branes and anti-branes at the fixed points. Take the $\IZ_3$
orientifold with an action of the twist on the CP factors given by
: \beq
V_9 \ = \  1/3(1,1,1,-1,-1,0,0,0,0,1,1,1,1,1,1,1) \eeq A
quantized Wilson line is added in the first complex plane given
by:
\beq
W_9 \ = \  1/3(1,1,1,1,1,-1,0,0,0,0,0,0,0,0,0,0) \eeq The
gauge group from the (99) sector will be $U(3)\times U(2)\times
U(1)\times [SO(6)\times U(7)]$. Now, the 9  $\IZ_3$ fixed points in
the first two complex planes split into three sets of three fixed
points each : $(0,m)$, $(1,m)$ and $(-1,m)$ with $m=0,\pm 1$. The
shifts corresponding to each of these three sets are $V$, $V+W$
and $V-W$ respectively. The corresponding value for $\Tr
\gamma_{\theta,9}$ are -4, -1 and 2 respectively. Looking at the
tadpole cancellation conditions we see that we will be forced to
add $5$-branes at the fixed points $(1,m)$ with $\Tr
\gamma_{\theta,5}= -1$ and also at the fixed points $(-1,m)$ with
$\Tr \gamma_{\theta,5}= -2$. No $5$-branes have to be added to the
points $(0,m)$ since tadpoles cancel without them. Thus the
simplest option is to locate 5-branes with twist actions given by
:
\beqa
(1,m)   \rightarrow  \  \gamma _{\theta,5}  & = & \diag  (
\alpha ,\alpha^2 ) \\ \nonumber (-1,m)   \rightarrow \  \gamma
_{\theta,5}  & = & \diag  (\alpha I_2,\alpha^2 I_2) \eeqa One gets
the following massless spectrum:

{\bf (99):}

There are  chiral multiplets  transforming under the gauge group
like (the hypercharge  $U(1)$ generator is as in section 4.1):
\beq
3(3,2,1/6)\ +\ 3({\bar 3},1,-2/3) \ +\ 3(1,2,+1/2) \ +\ [
3(21)' +3(6,{\bar 7})'] \eeq These fields  correspond to 3
generations of left-handed quarks and right-handed U-quarks as
well as three sets of Higgs fields $H_U$.

{\bf (59) :}

{}From the 5-branes at the $(1,m)$ fixed points we get a $U(1)_5^3$ gauge
group. One then finds  chiral multiplets (in three copies, one per
fixed point) transforming like:
\beq
({\bar 3},1,1/3)_q \ +\
(1,1,-1)_q \  +\ (1,2,-1/2)_{-q}\ +\  (1,2,1/2)_{-q}
 \ +
\ [ 7'_{q}\ +\ 6'_{-q}] \eeq where the subindex $q$ refers to the
$U(1)^3_5$ charges. These fields include three  generations of
right-handed $D$-quarks  and three sets of vector-like doublets .
The latter can become massive if the singlets in the $(55)$ sectors
get a vev.
{}From the 5-branes at the $(-1,m)$ fixed points we get a $U(2)^3$
gauge group. One then finds  chiral multiplets (in three copies,
one per fixed point) transforming like:
\beqa
(1,2,-1/2;2)_r  \ &
+&  \ (1,1,+1;2)_r \ +\
 ({\bar 3},1,1/3;2)_{-r} \  + \ (3,1,-1/3;2)_{-r} \\ \nonumber
  &  + &  \ [ ( 7';2)_r \ +\ 6'_{-r}]
\label{aqui} \eeqa where the subindex $r$ refers to the $U(1)^3_5$
charges. This sector includes all leptons and additional weakly
interacting particles (as well as vector-like sets of extra color
triplets).
Now, the $(55)$ sector living on the $(-1,m)$ fixed points
contain an adjoint under the $SU(2)$'s which can get a vev
along a flat direction and give a mass to the
triplet anti-triplet pairs in the corresponding $(59)$
sector. Altogether this model contains
at the massless level three standard
quark-lepton generations plus three sets of vector-like
right- and left-handed extra leptons.

As we did in section 4.2, one can construct a related model with
both $5$- and anti-$5$-branes all stuck at the fixed points. The
simplest possibility is to replace the $5$-branes at the $(-1,m)$
fixed points  by anti-$5$-branes. We put two anti-$5$-branes with
$\gamma _{\theta, {\bar 5}}=(I_2)$. This meets the tadpole
conditions and leads to a group $Sp(2)^3$. Now, the spectrum of
this model is similar to the previous one but replacing the SUSY
spectrum in (\ref{aqui}) by the following (non-SUSY) one:

{\bf (${\bar 5}9$) : }

\beqa
Fermions &:& \ \  (1,2,-1/2;2)  \  +  \ (1,1,+1;2) \  +\ [
(7';2)  ] \\ \nonumber Scalars  &:& \ \ ({\bar 3},1,1/3;2) \ + \ (
3,1,-1/3;2)\ +\ [(6';2)] \eeqa
 As in the previous situation, this
sector provides for the leptons which then are the only non-SUSY
sector of the model. The ${\bar 5}{\bar 5}$ sector contains only
one negative chirality triplet for each of the $Sp(2)$ groups.

\subsection{Another Variant SM}

Here we will just write the spectrum of the other  model discussed
at the beginning of section 4 containing $SU(3)\times SU(2)\times
U(1)$. $N_2=9, N_0=1$; $n_1^i=m_1^i=m_0^i=1$. The full gauge group
is
\beq
[SU(3)\times SU(2)\times U(1)]\times U(1)^3\times [SO(2)\times
U(9)]' \times Sp(2)^6\times U(1)^3 \eeq
 The hypercharge generator can be defined as in the previous
models. The
spectrum is

{ 99}  sector:

\beq
   3\, [(3,2,1/6) + (1,2,1/2) + (\bar 3, 1,-2/3)] + 3\ [(2,9)' + (1,36)']
\eeq
Again this sector can incorporate the three families of left handed
quarks
and right handed $U$-quarks together with three sets of Higgs fields
$H_U$.

{}From the 5-branes at the points $L_0$ we have:

${ 5_{L_0}9}$  sector:

\beq
3\ [(\bar 3, 1,1/3; 2) + (1,2,-1/2;2) +  (1,9;2)'] \eeq

Where the index after the semicolon is the $Sp(2)$ representation
(one different for each fixed point). Here we have candidate leptons
and
right
handed $D$-quarks, although they are doublets of $Sp(2)$ also.
{}From the 5-branes at the points $L_1$ the massless spectrum  is:

${ 5_{L_1}9}$  sector:

\beqa
3\ [ (\bar 3,1,1/3)_q + (1,2,-1/2)_{-q} + (1,2,1/2)_q + (1,1,-1)_q +
(2,1)_{-q}'+(1,9)'_{q}]\nonumber\\ + 3\ [(3, 1,-1/3; 2) + (1,1,-1;2)]
+ (1,\bar 9;2)']
\eeqa

Where the subindices $q$ are the charges with respect to each of the
three $U(1)$'s coming from the 5-branes. Here we also have candidate
leptons and right handed $D$-quarks.

Finally, the massless spectrum of the 55 sector transform under
each $Sp(2)\times U(1)$ at the $L_1$ fixed points as:
${ 55_{L_1}}$  sector:

\beq
3\ [(2)_{-q}+ 2\ (1)_{2q} + (1)_{2q}] \eeq

Contrary to the previous models, it is not clear how in this model all
the
extra (unwanted) triplets and doublets can get a mass, therefore this
model looks at  first sight less interesting phenomenologically,
although
a detailed study may reveal ways of making it more realistic.
A similar model can be obtained by substituting the branes at the
${L_1}$ points by four anti branes ($q_1^i=1$). Therefore the
spectrum is the same except for the ${\bf 5_{L_1}9}$ sector
above which is changed by the nonsupersymmetric one (with  opposite
chirality)
in three copies:

\beqa
Fermions &:& \ \  (\bar 3,1,1/3)_r\ + \ (1,2,1/2)_r  \  +  \
(1,2,-1/2)_{-r}\
+ \ (1,1,+1)_r \ \nonumber \\
& &  +\ [(2,1)_{-r}'\ + \ (1,9)_r']  \nonumber\\
 Scalars  &:& \ \ ({ 3},1,-1/3)_{-r} \ + \ (
1,1,+1)_r\ +\ [(1,9)_{-r}'] \ + \ h.c. \eeqa

Where $r$ is the charge with respect to each of the $U(1)$'s coming
from the anti branes.
 The $\bar 5\bar 5_{L_1}$ also replaces the $55_{L_1}$ sector which
consists of three fermions and six scalars (singlets) with charge
$-2r$.
It also includes three extra singlets with negative chirality.
Therefore, we may have here both
leptons and right handed $D$-quarks in a non-supersymmetric sector.

\subsection {Continuous Wilson lines. Models with adjoint representations}

As we have stressed, in contrast to what happens in supersymmetric
models, the addition of antibranes allows for several consistent
twist matrices and Wilson lines.  In particular continuous Wilson
lines can be easily included. This is an interesting possibility
for model building since it can be used for reducing the gauge
groups ranks and for obtaining adjoint representations. Here we
show  how continuous Wilson lines can be introduced  in $\IZ_3$
orientifold discussed in section 3. Let us consider the case
$N_1=N_3=N_4=N$, and thus $N_0+N_2=16-N$ in \ref{gtilde} and
rewrite the twist matrix and discrete Wilson line ${\cal W}$ in terms of
$3N\times 3N$ and $(16-3N)\times (16-3N)$ matrices as ${\tilde
{\gamma }}=({\gamma }_{diag},{\gamma }_2)$ where
\beqa
{\gamma
_{diag}}  & = & \diag  (I_{N},\alpha I_{N},\alpha^2 I_{N}) \\
{\gamma _2}  & = & \diag  (I_{N_0},\alpha I_{N_2}) \eeqa

and
\begin{equation}
  {\tilde {\cal W}} =  \diag
(\alpha I_{N},\alpha I_{N}, \alpha I_{N}; I_{N_0},I_{N_2})
\end{equation}

Hence, the gauge group $U(N)^3 \times SO(2N_0)\times U(N_2)$ is
obtained, and the massless spectrum can be read from \ref{99spec}
to be \footnote{ There we must perform the  replacement ${\ov {\bf
N}}_4 \rightarrow
 {\bf N}_4$, due to the term $\alpha^2 I_{N_4}$ in the twist matrix
 above}

\begin{eqnarray}
& &  3[({\ov {\bf N}},{\bf N},1;1,1)+ (1,{\ov {\bf N}},{\bf
N},1;1,1)+ ({\bf N},1,{\ov {\bf N}};1,1)]+ \\ & &
+3[(1,1,1;2N_0,N_2) +(1,1,1;1,{\bf a}_2)] \label{z12spec}
\end{eqnarray}

 which is invariant under the permutations of the $U(N)$ gauge
 groups. Moreover, by giving appropriate $vev$'s to
 $(\ov {\bf N},{\bf N})$ multiplets, $U(N)^3$ group should break down
 to
 the diagonal
subgroup $SU(N)_{diag}\times U(1)$, leaving three adjoint
representations in the spectrum.

In particular an $SU(5)$,  group with three $\bf {24}$ massless
representations could be obtainable. However, since antisymmetric
representations of the ${\bf 99}$ gauge group cannot appear in
crossed sectors, the model is expected to e non chiral.

Such a breaking is easily achieved by considering a continuous
Wilson line of the type presented in \cite{afiv}. Namely,

\begin{equation}
{\tilde {\cal W}}=  \left ( \begin{array}{cc}
 W_{diag} & 0  \\
0 & 1_{N_0+N_2} \\
\end{array}
\right )
\end{equation}
a $16 \times 16 $ matrix, where  $W_{diag}$ is a $3N\times 3N$
matrix (which acts on ${\gamma }_{diag}$), defined as
\begin{equation}
W_{diag}=  \left ( \begin{array}{ccc} \lambda & a & a \\ a &
\lambda & a\\ a & a & \lambda
\end{array}
\right )
\end{equation}
where $a$ and $\lambda$ are  arbitrary complex numbers and each
block is a $n\times n$ matrix. It is straightforward to check that
$ ( \gamma _{diag} W_{diag})^3=cI_{3n} $  where
\begin{equation}
c= (\lambda ^3-3a^2 \lambda +2a^3)
\end{equation}
and also $\Tr ( \gamma _{diag} W^k_{diag})=0$ $k=0,1,2$. Therefore
\beq \Tr  \gamma _{9} ({\cal W})^k= 2N_0-N_2 \label{trgw} \eeq for
$k=0,1,2$.

Thus, we see  that for $a \ne 0$ (and $\lambda/a\ne 1,-2$)  $a$
can be chosen such that $c=1$. Hence, we are still left with a
complex continuous parameter defining a continuous Wilson line
 \cite{inq2,afiv}.

Performing the usual projections, such a continuous Wilson line
achieves the desired breaking to the diagonal group by keeping
three adjoints in the ${\bf 99}$ spectrum. As mentioned, chiral
generations do not appear in this sector.

{}From (\ref{trgw})
we  see that the contribution to tadpole equation
(\ref{tadpzz}) comes only from the non  rotated piece. Therefore, we
have
\beq 2N_0-N_2 + 3(\Tr\gamma_{\theta,5,L}-
\Tr\gamma_{\theta,\bar{5},L}) = -4
\eeq
with $N_0+N_2= 16-3v$. This constraint can be satisfied, for
instance, by placing five branes at each of the nine $\IZ_3$ fixed
points, satisfying $\Tr\gamma_{\theta,5,L}=\frac{1}3(-4+2N_0-N_2)=
-12+2N+N_2$ and the same number of antibranes in the bulk. Other
choices, with branes and antibranes stuck a these points are also
possible.

For instance, for $N=5$ and choosing $N_2=1$, tadpole equations
are satisfied by placing two five branes with
$\gamma_{\theta,\bar{5},L}= (\alpha I_{1},\alpha ^2 I_{1})$ (thus
leading to a $U(1)$ group at each ${\bf 55}_{L_a^i}$ sector) and
eighteen antibranes in the bulk.

This choice leads to $(SU(5)\times U(1))_{diag}\times U(1)$ gauge
group with  three $\quad {\bf {24}}$ adjoint representations in
the ${\bf 99}$ sector. We also have ${\ov {\bf 5}}_{1}+{\bf 5}_1$
and ${\ov {\bf 5}}_{-1}+{\bf 5}_{-1}$ non chiral combinations at
each of the nine fixed points, where the subindex indicates the
charge with respect to the $U(1)_ {55}$ group at each
point.

\end{document}